\documentclass{article}

\usepackage{subfigure}

\usepackage[utf8]{inputenc} 
\usepackage[T1]{fontenc}    
\usepackage{hyperref}       
\usepackage{url}            
\usepackage{booktabs}       
\usepackage{amsfonts}       
\usepackage{nicefrac}       
\usepackage{microtype}      
\usepackage{xcolor}         
\usepackage{enumitem}

\usepackage{amsmath,amsfonts}
\usepackage{algorithmic}
\usepackage{graphicx}
\usepackage{textcomp}
\usepackage{multirow}
\usepackage{xcolor}
\usepackage{circledsteps}
\usepackage[ruled,linesnumbered]{algorithm2e}
\usepackage{soul}
\usepackage{layouts}
\usepackage{xcolor}
\usepackage{tcolorbox}
\usepackage{soul}
\usepackage{url}
\usepackage{hyperref}
\usepackage{censor}
\usepackage{scrextend}
\usepackage{tcolorbox}
\tcbuselibrary{skins} 
\usepackage{hyperref}

\NewTColorBox {calloutbox} { O{} O{} } 
{
  title={#1},
  coltitle=white,           
  colframe=blue,            
  colback=blue!10!white,    
  sharp corners,            
  #2                        
}

\usepackage[accepted]{mlsys2025}

\newcommand{\name}[1]{\textit{LIMINAL}}

\newcounter{tcount}
\newcommand{\takeawayremoved}[1]{}

\newcommand{\trimmed}[1]{}

\mlsystitlerunning{The Impact Market to Save Conference Peer Review: Decoupling Dissemination and Credentialing}

\begin{document}

\twocolumn[
\mlsystitle{The Impact Market to Save Conference Peer Review: Decoupling Dissemination and Credentialing}



\mlsyssetsymbol{equal}{*}

\begin{mlsysauthorlist}
\mlsysauthor{Karthikeyan Sankaralingam}{uw}
\end{mlsysauthorlist}

\mlsysaffiliation{uw}{UW Madison}
\mlsyscorrespondingauthor{Karthikeyan Sankaralingam}{karu@cs.wisc.edu}

\mlsyskeywords{Machine Learning, MLSys}
\pagestyle{fancy}
\pagenumbering{arabic}
\vskip 0.3in

\begin{abstract}
Top-tier academic conferences are failing under the strain of two irreconcilable roles: (1) rapid dissemination of all sound research and (2) scarce credentialing for prestige and career advancement. This conflict has created a "reviewer roulette" and "anonymous tribunal" model—a "zero-cost attack" system—characterized by high-stakes subjectivity, "turf wars," and the arbitrary rejection of sound research (the "equivalence class" problem).
We propose the Impact Market (IM), a novel three-phase system that decouples publication from prestige.
\textbf{Phase 1 (Publication)}: All sound and rigorous papers are accepted via a PC review, solving the "equivalence class" problem.
\textbf{Phase 2 (Investment): }An immediate, scarce prestige signal is created via a "futures market." Senior community members "invest" tokens into published papers, creating a transparent, crowdsourced Net Invested Score (NIS).
\textbf{Phase 3 (Calibration):} A 3-year lookback mechanism validates these investments against a manipulation-resistant Multi-Vector Impact Score (MVIS). This MVIS adjusts each investor's future influence (their "Investor Rating"), imposing a quantifiable cost on bad actors and rewarding accurate speculation.
The IM model replaces a hidden, "zero-cost attack" system with a transparent, accountable, and data-driven market that aligns immediate credentialing with long-term, validated impact. Agent-based simulations demonstrate that while a passive market matches current protocols in low-skill environments, introducing investor agency and conviction betting increases the retrieval of high-impact papers from 28\% to over 85\% under identical conditions, confirming that incentivized self-selection is the mechanism required to scale peer review.
\end{abstract}
]



\printAffiliationsAndNotice{} 

\section{Introduction}
\begin{addmargin}[0.1in]{0.1in}
\textit{We are told by the senior guardians of our field—all honorable people—that our system of review is sound.
}

\textit{We are told that the 15\% acceptance rate is the necessary cost of ``prestige,'' and that the 40\% ``equivalence class'' of sound-but-rejected papers is a `necessary sacrifice' for this greater good. Our system works, we are told. And our guardians are honorable people.}

\textit{We are told that the ``reviewer roulette'' of the ``anonymous tribunal'' is rigorous debate, that the ``nonsensical'' review is an unfortunate anomaly, and that the ``zero-cost attacks'' are not a feature, but a bug. And they are, all, honorable men.}

\textit{I do not come to speak in praise of this system. I come to bury it.}

\textit{For the system they defend has, in its quest for ``prestige,'' murdered ``merit.'' And I am here to read the will\footnote{If you haven't already gathered, this is adapted from~\cite{juliusceasar}}.}
\end{addmargin}
\medskip

The top-tier academic conference is a cornerstone of computer science, a multi-faceted institution that has served the community for decades. It simultaneously serves as a social gathering for researchers, an intellectual forum for the presentation of new ideas, and, critically, a primary mechanism for credentialing. Acceptance at a ``top conference'' is the signal used for hiring, tenure, awards, and funding. This single, scarce signal of prestige is, for many, the conference's most important function.

At this point, some may argue that open-access repositories like ArXiv have already solved the dissemination problem. This argument, however, fails to address the core conflict. ArXiv provides rapid dissemination but offers zero credentialing. Its model for impact is outsourced entirely to informal, external mechanisms: an author's pre-existing reputation, their institutional prestige, or their savvy in promoting work on social media platforms~\cite{Bagchi2025Effects,Weissburg2024Tweets}. This ``ArXiv-plus-social-media'' model actively penalizes researchers who, for valid personal or ethical reasons (such as concerns over rampant misinformation, malice, and political bias), choose not to engage with platforms like X or Facebook. Thus, ArXiv does not solve the credentialing crisis; it simply displaces it onto a less structured and equally arbitrary stage. The top-tier conference, therefore, remains the central battlefield for prestige.

However, this system is fundamentally broken. It was designed for a 1980s-level of research output and can no longer cope with the volume, diversity, and incentives of the 21st-century academic world. The result is a system in crisis, characterized by three core failure modes:

\textbf{The ``Equivalence Class'' Problem:} As fields like computer architecture have matured, a large percentage of submissions (informally estimated at 40-50\%) fall into an ``equivalence class'' of high-quality research. These papers are sound, experimentally rigorous, and novel. Yet, due to arbitrarily low acceptance rates (a requirement for ``prestige''), the vast majority are rejected. This is no longer meritocracy; it is a lottery. For junior authors, not used to some of the historical literature, our observations on failures on the peer review model are not novel or surprising. Skip ahead to the related work. The two seminal works are: i) Thurner and Hanel \cite{Thurner2011} provided the game-theoretic foundation for peer-review, demonstrating that rational reviewers in a competitive environment are incentivized to reject excellent papers that threaten their own status (``turf wars''). ii) Price~\cite{Price2014,cortes2021inconsistencyconferencepeerreview} did the famous experiment in NEURIPS 2014, showing that a subset of papers when reviewed by two disjoint PC had a disagrrement of 57\%, implying the ``true'' acceptance rate is much higher. More recently, a similar study was done for NEURIPS 2021~\cite{neurips2021inconsistency}. That paper concluded: \textit{``Finally, we would encourage authors to avoid excessive discouragement from rejections as there is a real possibility that the result says more about the review process than the paper.''} For a PhD Supervisor or Research Lab lead, this arbitrariness is cost-free, and the accepts cancel out the rejects. Which is arguably why the collective leadership has allowed this broken system to continue so long. For an individual early stage researcher though, these rejects define their career trajectories. 

One might then ask a simple question: why not just increase acceptance rates to 50\% and solve the problem? This response, however, ignores the political reality of the academic ecosystem. Such a move would unilaterally destroy the conference's second, and arguably most dominant, function: credentialing. Tenure committees, hiring managers, and funding agencies depend on the scarcity of this signal. If the signal vanishes, they will simply invent new, less transparent proxies for prestige—such as institutional pedigree or personal networks—which would only amplify the system's existing biases. The problem is not that credentialing exists, but that our method of credentialing has become a lottery that blocks dissemination. The two functions, namely dissmenination of sound research and credentialling, must be decoupled, not destroyed.

\textbf{The ``Reviewer Roulette'':} The binary, high-stakes nature of review has devolved into an ``anonymous tribunal.'' Because a paper's soundness is often not in question, reviewers are forced to litigate its significance. This subjective criterion is a weapon for ``turf wars'' and a hiding place for reviewer incompetence, allowing non-scientific factors to dominate scientific review. Cortes and Lawrence~\cite{cortes2021inconsistencyconferencepeerreview} showed there is essentially \textit{no correlation between the review scores and impact} (measured by number of citations with a log transformation) - meaning the very cudgel, ``significance'' used to reject and rate papers is not correlated with eventual impact. 

\begin{calloutbox}
The central thesis of this paper is that these problems are symptoms of a single, fatal design flaw: we are using a blunt, binary instrument (Accept/Reject) to perform two fundamentally distinct jobs: (1) Dissemination and (2) Credentialing. This system can no longer do both.
\end{calloutbox}

We propose a solution that decouples these signals. This paper introduces the Impact Market (IM), a novel, three-phase system that replaces our broken review model. The (IM) is designed to:

\begin{itemize}
\item \textbf{Accept all sound papers for dissemination.} It solves the ``equivalence class'' problem by moving the Program Committee's job from ``gatekeeper'' to ``validator.''

\item \textbf{Create a new, parallel signal for prestige.} It creates a transparent, crowdsourced ``futures market'' where experts (the program committee and external review committee pool of the conference) invest tokens in papers they believe will have a long-term impact. This generates an immediate, scarce Net Invested Score (NIS) for credentialing.

\item \textbf{Be self-correcting and data-driven.} It uses a 3-year lookback to validate these speculations against a manipulation-resistant ``impact score.'' This feedback loop punishes bad-faith collusion and marginalizes incompetent reviewers, ensuring the system's long-term health.
\end{itemize}

Through agent-based simulation, we quantify the ``cost of noise'' in the current system, showing that it functions indistinguishably from a lottery for high-impact work. In contrast, we show that by incentivizing agency and conviction, the Impact Market recovers over 85\% of \textbf{gem} papers under identical conditions, confirming that market dynamics are essential for scaling evaluation. When the expert ability of the peer group is tuned to a beta(5,1) function, recall reaches almost 100\%.

This paper will first detail the critical failures of the current system (\S\ref{sec:current}). It will then present the full architecture of the Impact Market (\S\ref{sec:design}). Finally, it will analyze the IM's resilience to the precise problems—collusion, incompetence, and turf wars—that plague our community today (\S\ref{sec:analysis}), including a roadmap for incremental adoption and transition to the IM. We discuss a simulation methodology to study the protocol (\S\ref{sec:methodology}), and results from simulation in~\S\ref{sec:results}. Related work is discussed in~\S\ref{sec:related}. The paper includes a detailed discussion of limitations in~\S\ref{sec:limitations} and five detailed appendices. Appendix~\ref{ref:formalprotocol} includes a formal specification of the Impact Market and analysis of its capabilities. Appendix~\ref{ref:eqclass_appendix} provides substantiation for the higher acceptance rates argued here. Appendix~\ref{ref:transparency_appendix} describes the implications of transparency and how it might conflict with the goals of so-called arms-length letters. Appendix~\ref{ref:goldenage_appendix} analyses the ``golden age'' of peer review (80s and 90s), arguing that its success was based on an unscalable protocol of social accountability—a system whose collapse was not just predictable, but inevitable. Appendix~\ref{sec:collusion_appendix} discusses collusion detection, Appendix~\ref{ref:longtail_appendix} details data on citation counts of published papers showing the long tail of impact. Finally Appendix~\ref{sec:senior_appendix} develops in depth resistance from established researchers for whom the current system works. The source code for the simulator is available here: \url{https://github.com/VerticalResearchGroup/ImpactMarket}.

This work makes the following contributions:
\begin{itemize}
    
\item \textbf{Reframing Peer Review as a Protocol-Level Incentive Failure.}
We identify that the dominant conference peer-review model couples validation and credentialing into a single binary decision, creating a zero-cost attack surface for biased, incompetent, or strategically harmful reviewing. We formalize this as a protocol-design flaw rather than a cultural or resourcing problem.

\item \textbf{A Three-Phase Evaluation Architecture that Decouples Dissemination from Prestige.}
We introduce the Impact Market (IM), consisting of (i) Publication based solely on soundness, (ii) Investment using bounded reputational tokens to generate a predictive signal of impact, and (iii) Calibration against a Multi-Vector Impact Score (MVIS) to update Investor Ratings (IR). This separation removes scarcity from dissemination while retaining a prestige signal aligned with long-term impact.

\item \textbf{A Self-Calibrating Reputational Mechanism with Formal Incentive Analysis.}
We model the IM as a dynamic system and prove that (a) investors with high IR have increasing influence only when they correctly identify impactful work, (b) misaligned or noisy investors lose influence over successive calibration cycles, and (c) attack strategies—including popularity cascades and collusive amplification—become high-cost and self-limiting.

\item \textbf{Simulation Evidence of Convergence and Attack Resistance.}
Through agent-based simulations, we demonstrate that (i) Net Invested Score (NIS) converges toward MVIS-aligned rankings over iterations, (ii) collusive behavior collapses when MVIS feedback penalizes coordinated misprediction, and (iii) controversial papers with high true impact surface earlier than under binary peer review.

\item \textbf{A Practical Deployment Pathway.}
We outline a low-risk shadow-deployment protocol for running IM alongside existing conference review processes, enabling empirical validation without disrupting publication pipelines, and establishing a feasible path toward adoption.

\end{itemize}

We acknowledge that this proposal is disruptive and will invite immediate, cynical skepticism. Objections will certainly arise: that a ``market'' will devolve into a ``popularity contest,'' that ``investments'' will be gamed by collusion, or that the calibration protocol is simply too complex. As we will demonstrate, the IM has been designed from the ground up with these precise attacks in mind. It incorporates a Two-Bucket Budget to solve the popularity problem, a MVIS-feedback-loop to make collusion a provably self-destructive strategy, and a workflow that, for the average expert, is less work than the current model. Some will even argue the system works, simply denying the data staring us in our face: the famous NEURIPS 2014 and 2021 experiment are the gold standard, showing disagreement rate is shockingly 57\%. Or defend current peer review as ``subjectivity'' is important. To these readers, we encourage reading this paper with an open mind and address those concerns in~\S\ref{subsec:systemworks}.

The Impact Market has the hallmark of a robust solution: it is not a minor tweak to a broken process, but a new architecture built on a foundation of transparency, data-driven accountability, and explicit alignment of incentives. It is also not a utopian fantasy. The tools to build this system—reputation protocols, multi-vector metrics, and automated feedback loops—are the very tools our field creates and understands. This is a system that can be imagined, and as we will show, it is a system that can be put in play today.

\section{Preliminaries and Framing}

\subsection{Scope of This Proposal}

This proposal is not intended as a universal panacea for all of academia. It is explicitly designed for communities, primarily in Computer Science and Engineering (e.g., SIGARCH, SIGOPS, SIGPLAN, ASPLOS, ISCA, OSDI, SIGCOMM), that share a specific set of cultural and structural properties:

\begin{itemize}
\item A conference-centric (not journal-centric) publication model.
\item A high volume of submissions where soundness is often verifiable, but significance is the main point of contention.
\item A strong ``artifact'' culture (code, datasets) that makes a robust ``Multi-Vector Impact Score'' (MVIS) feasible.
\end{itemize}

The IM protocol is a template for these communities. Other fields, such as theoretical mathematics or the humanities, would likely require a different set of protocols.

\subsection{Fundamental assumptions}

This proposal rests on three foundational assumptions that some readers may dispute:

\textbf{Assumption 1: Impact can be defined.} We assume there exists a meaningful concept of "scientific impact"—that some research contributions matter more than others for advancing knowledge, enabling applications, or influencing the field's direction.

\textbf{Assumption 2: Impact can be measured quantitatively and publicly.} We assume that while no single metric perfectly captures impact, a composite of observable signals (citations, artifact adoption, cross-disciplinary influence) provides a reasonable approximation that can be measured and validated over time. The view that impact is the subjective opinion of leaders in the field we simply summarily dismiss.

\textbf{Assumption 3: Prestige should align with impact.} We assume that it is desirable for academic credentialing systems to reward research based on its demonstrated contribution to science, rather than on luck, social networks, or institutional affiliation.

Readers who reject these assumptions will naturally reject the IM. However, we note that \textbf{the Current Protocol implicitly makes identical assumptions}—it simply implements them catastrophically poorly. When reviewers reject papers for "insufficient significance," they are asserting that impact can be defined and judged. When tenure committees weight top-venue publications, they are asserting that prestige should reflect quality. When hiring committees informally check citation counts, they are asserting that impact can be measured. The CP does not avoid these assumptions; it executes them through opaque, unaccountable, and arbitrary mechanisms. The IM makes the same assumptions explicit and implements them transparently with data-driven validation.

\subsection{The Reality of Scientific Impact: A Long-Tail Distribution.}
\begin{figure}[t]
\centering
\includegraphics[width=\columnwidth]{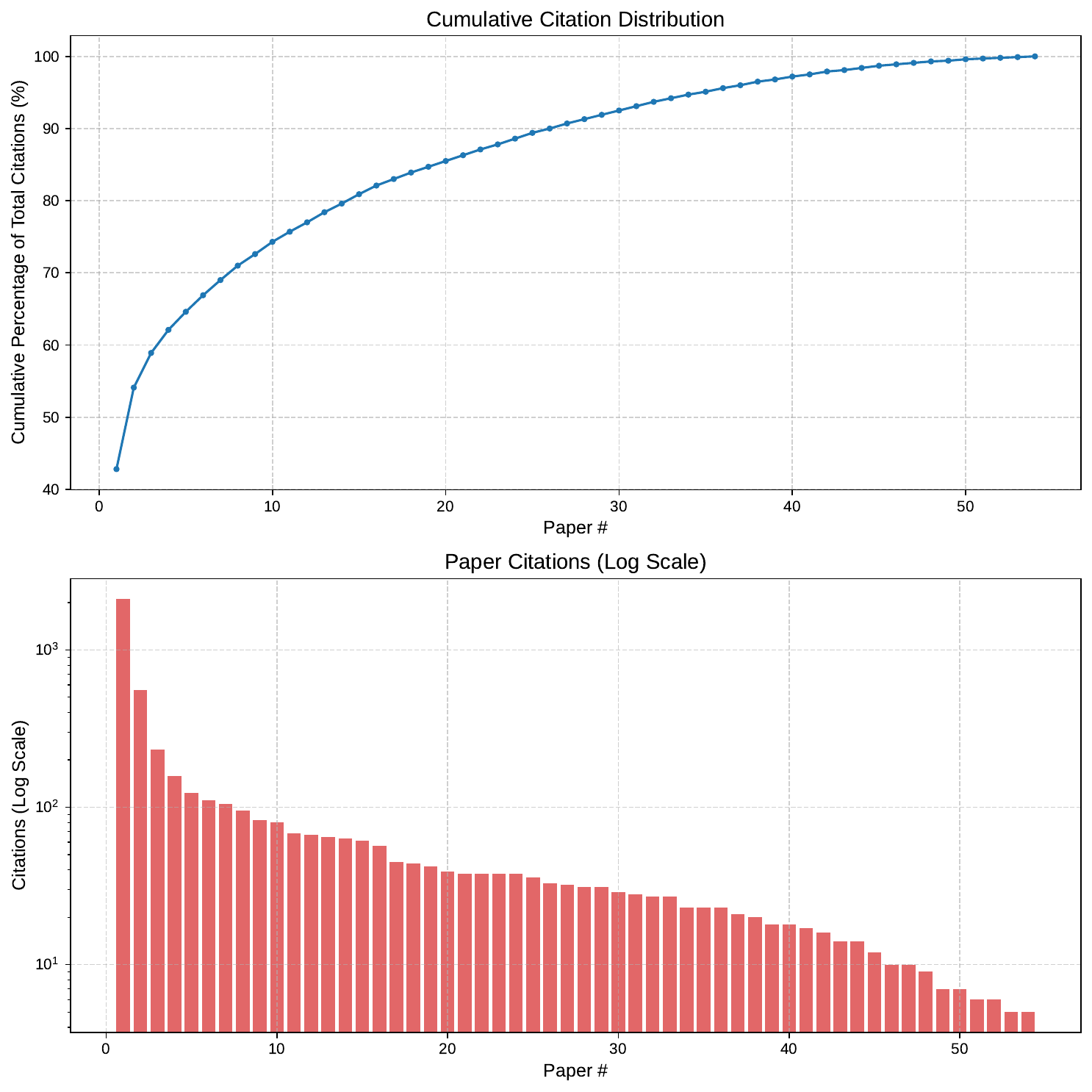}
\caption{\textbf{Citation distribution for ISCA 2017 (55 papers, 7 years post-publication).} Top: Cumulative percentage of total citations. Bottom: Individual paper citation counts (log scale).}
\label{fig:isca2017}
\end{figure}

Scientific impact is not uniformly distributed—it follows a steep power law. Figure~\ref{fig:isca2017} shows citation data for ISCA 2017 originally collected by~\cite{10.7717/peerj-cs.1389}, seven years post-publication. The top 10 papers (18\% of the conference) account for 75\% of all citations from papers published in ISCA 2017. The top 20 papers (36\%) account for 86\%. The remaining 35 papers—nearly two-thirds of accepted work—share only 14\% of citations. Table~\ref{tab:citations_tables} shows this data for another 10 systems conferences. Appendix E presents full distributions and more commentary.

\begin{table*}[tbp]
    \centering
    \tiny
    \begin{tabular}{c|c|c|c|c|c|c|c|c|c|c|c}
Papers          &ISCA           &ASPLOS         &SIGCOMM        &PLDI           &OOPSLA         &MICRO          &SOSP           &NSDI           &MobiCom        &SC             &\\ \hline
& \multicolumn{11}{c}{For every conference a triad of \% of papers 10 or 20 papers accounts for, \% of citations covered, average citations} \\ \hline
Top 10          & (19/74/364/)   &(18/52/118/)   &(28/59/202/)   &(21/57/97/)    &(15/47/80/)    &(16/50/105/)   &(26/68/268/)   &(24/53/168/)   &(29/63/102/)   &(16/47/79/)    \\
Top 20          & (37/86/210/)   &(36/71/81/)    &(56/84/142/)   &(43/79/67/)    &(30/66/56/)    &(33/70/74/)    &(51/87/170/)   &(48/78/123/)   &(57/88/71/)    &(33/67/57/)    \\ \hline
& \multicolumn{11}{c}{For every conference a triad of \# of papers 80\%, 75\%, 50\%,  papers accounts for, \% of citations covered, average citations} \\ \hline
Bot-80\%        & (43/24/28/)    &(44/43/22/)    &(28/47/57/)    &(37/43/20/)    &(52/44/15/)    &(48/43/19/)    &(31/37/47/)    &(33/50/48/)    &(28/51/29/)    &(48/46/16/)    \\
Bot-75\%        & (40/20/25/)    &(42/39/21/)    &(27/44/55/)    &(35/38/19/)    &(49/39/14/)    &(45/37/17/)    &(29/32/43/)    &(31/44/45/)    &(26/42/26/)    &(45/40/15/)    \\
Bot-50\%        & (27/9/17/)     &(28/17/14/)    &(18/20/38/)    &(23/16/12/)    &(33/17/9/)     &(30/15/11/)    &(19/13/28/)    &(21/20/31/)    &(17/16/15/)    &(30/18/10/)  \\ \hline
    \end{tabular}
    \caption{Citations for many top-tier conference in 2017, four years after publications. The Top-10 and Top-20 rows refer to data for the top-10 and top-20 most cited papers. The three numbers in the cells refer to the \% total papers those papers account for, \% of total citations they account for, average citations for that group. For the Bot-80\%, Bot\-75\% and Bot-50\% refers to least cited papers covering the lowest 80\%, 75\%, and 50\%. For those the 3 numbers refer to the number of papers in that class, percentage of total citations in the conference they cover, and average \# citations.}
    \label{tab:citations_tables}
\end{table*}

This is the reality of science~\cite{10.7717/peerj-cs.1389, 10.1145/1531793.1531815}. Most papers, even at top venues, have modest impact. The Current Protocol obscures this by forcing binary Accept/Reject decisions. The Impact Market embraces it: Phase 2 predicts a continuous NIS distribution matching the long-tail structure, and Phase 3 validates accuracy against actual impact.

A common concern: ``Won't low NIS be discouraging?'' The data answers this. If 70\% of ISCA 2017 papers have fewer than 30 citations 4 years after publication, then 70\% of papers \textit{should} have modest NIS\footnote{The issues of sub-areas having their own citation volumes and dynamics will simply be translated to NIS - this is not unique to the IM or NIS}. The IM is honest about this reality; \textit{the current peer review process lies about it by rejecting half of all sound papers to maintain artificial scarcity.}

\subsection{What Is Outside Scope.}

Several important concerns are orthogonal to this work.

\textbf{Confidential or Non-Observable Impact:} Some might argue that certain research has profound impact that cannot be publicly measured—e.g., private consultations with companies, classified government work, or ideas communicated through informal channels. We acknowledge such cases exist but argue they are rare and irrelevant to the problems we address. Work whose impact is inherently non-observable cannot be credentialed by \textit{any} public peer review system, including the CP. Researchers doing such work are already operating outside the conference publication model. The IM is designed for the 99\% of research whose impact \textit{is} eventually observable through standard scholarly channels.

\textbf{Whether Prestige Should Exist at All:} In an ideal world, science might operate without prestige, competition, or credentialing—researchers would pursue knowledge purely for intrinsic motivation, and society would fund science without requiring individual-level evaluation. This utopian vision is outside our scope. We operate in a world where tenure committees, funding agencies, and hiring managers \textit{do} require prestige signals, and we propose a system that makes those signals more accurate and fair. Eliminating prestige from science is a separate (and much harder) problem.

\textbf{Aligning Science with Societal Good:} The IM measures impact within the scientific community (citations, adoption, influence) but does not address whether that impact serves humanity's interests. A paper enabling surveillance technology or a model trained on ethically problematic data might have high citations and thus high MVIS, even if its societal impact is net negative. Designing incentive structures that align scientific progress with human flourishing is a profound challenge that extends far beyond peer review reform. The IM does not solve this; it solves the narrower problem of making credentialing reflect scientific impact rather than reviewer lottery. Addressing the broader question—"What science should we incentivize?"—requires changes to funding priorities, ethical review processes, and societal governance of technology. These are critical issues, but they are orthogonal to fixing the mechanical failures of conference peer review.

\textit{We focus narrowly on a tractable problem: the Current Protocol is failing to disseminate sound research and is awarding prestige arbitrarily. The IM solves this by decoupling dissemination from credentialing and aligning prestige with measurable impact. This is not a complete solution to all problems in science, but it is a necessary foundation. A broken publication system that rejects half of all sound papers and rewards luck over merit cannot serve any larger goal—scientific, ethical, or societal. We must fix the foundation before we can build higher.}

\section{The Failures of the Current Peer Review Model}\label{sec:current}

The cognitive dissonance of the modern conference system—simultaneously a lottery and a high-stakes arbiter of careers—stems from a series of structural failures. The system is not just stressed; it is actively incentivizing the wrong behaviors and producing arbitrary outcomes. In the rest of this paper, we will refer to existing peer review as the ``Current Protocol (CP)''.

\subsection{The ``Equivalence Class'' Lottery}

As research fields mature, they become more successful, producing a larger volume of high-quality work. In fields like computer science and computer architecture, we have reached a point where a significant percentage of submissions, estimated at 40-50\%, fall into an ``equivalence class'' of sound research. These papers are (A) methodologically sound, (B) sufficiently novel, and (C) experimentally rigorous.

Under a healthy system of scientific dissemination, these papers would be published. Instead, they are forced into a brutal competition for a small number of ``prestige'' slots, which are kept arbitrarily low (e.g., 15-20\%) to maintain the signal of scarcity. The result is a lottery. A sound, valuable paper is just as likely to be rejected as accepted, with the final decision often hinging on the random draw of reviewers.

This 40-50\% estimate is not an arbitrary number. It is a data-driven heuristic derived from this author's empirical observations serving as a reviewer, meta-reviewer, and area chair for multiple top-tier systems conferences. This heuristic is based on a simple sampling: in a typical review stack of 15 papers, the Current Protocol (CP) accepts $\sim$15-20\% (e.g., 3 papers). However, in that same stack, another $\sim$30-40\% (e.g., 4-6 papers) are also sound, rigorous, and novel, but are rejected in the ``reviewer roulette.'' This combined group—the ``accepts'' plus the ``sound-but-rejects''—forms the ``equivalence class'' of papers that would pass the IM's Phase 1. This total is (15\% + 30\%) = 45\% to (20\% + 40\%) = 60\%. We use 40-50\% as a conservative estimate. The remaining submissions are the ``phantom group'' of truly flawed papers that would (and should) be rejected by any system for being demonstrably unsound or non-novel.

An alternative is the data from the NEURIPS 2014 experiment~\cite{cortes2021inconsistencyconferencepeerreview}, which allowed a subset of papers to be reviewed by multiple disparate groups. Their summary findings: ``the accept precision: if you are attending the conference and looking at any given paper, then
you might want to know the probability that the paper would have been rejected in an independent rerunning
of the conference. We can estimate this for Committee 1's conference as 22/(22 + 22) = 0.5 (50\%) and for
Committee 2's conference as 21/(22+21) = 0.49 (49\%). Averaging the two estimates gives us 49.5\%.''

The problems we identify—arbitrary outcomes, vulnerability to malice, and the ``equivalence class'' lottery—are not merely anecdotal. They have been rigorously documented and theoretically analyzed. Thurner and Hanel~\cite{Thurner2011} provide a foundational agent-based model demonstrating that peer review catastrophically fails when reviewers act rationally or selfishly rather than altruistically. Their model shows that even a small fraction of ``rational'' reviewers—those who have no incentive to promote high-quality work other than their own—is sufficient to drastically lower the quality of published work, to the point where peer review performs worse than pure chance. This is the basis of our core claim: the Current Protocol (CP) is a ``zero-cost attack'' system. Thurner and Hanel's work demonstrates that when reviewers face no accountability for biased or self-interested behavior, the system inevitably degrades.


\subsection{The ``Anonymous Tribunal'': A Crisis of Malice and Competence}

Because a binary ``Accept/Reject'' decision must be made, and a large portion of papers are sound, the review process is forced to litigate the one remaining subjective criterion: significance. This has devolved into an ``anonymous tribunal,'' a dysfunctional arbitration characterized by two key failures. The outcome is ``reviewer roulette.''

First is the Crisis of Malice. In an anonymous, high-stakes, binary-outcome system, reviewers are free to engage in non-scientific ``turf wars.'' The subjective ``significance'' vote becomes a silent veto, a perfect weapon for a rival to kill a competing paper without ever challenging its technical correctness.

Second, and more profoundly, is the Crisis of Competence. We must state plainly a truth that is widely observed by many in private but rarely said in public: a majority of our Program Committee (PC) and Extended Review Committee (ERC) members have no business doing peer review. This is not a failure of a few ``bad apples'' but a systemic failure of qualification.

This author's own experience as a meta-reviewer for recent top-tier conferences—where no personal papers were at stake—provides alarming evidence. When examining the reviews of PC members 4-8 years post-PhD, a disturbing pattern emerges. Reviews are frequently nonsensical, offering vacuous, catch-all criticisms like ``validate more'' or ``do more scalability'' without specific, actionable substance. In SIGARCH, when simulations are widely accepted, reviewers saying the weakness is the paper only does simulation. These are the comments of reviewers who lack the deep expertise to engage with a paper's core contribution but seemingly find a reason to reject it, to make it look like they are doing something thoughtful. Or worse, have learned from receiving mostly bad reviews, this behavior is what reviewing is about!

This widespread incompetence is the tinder for the ``reviewer roulette.'' It is often impossible to distinguish malice from simple ignorance. The subjective ``significance'' score is not only a weapon for the rival but also a hiding place for the unqualified reviewer. They can kill a paper they do not understand, not by engaging with its merits, but by simply declaring it ``not significant enough.''

\subsection{Why Existing Solutions Fail}

The community is not blind to these problems, but proposed solutions fail because they do not address the central conflict.

\textbf{Journal-First Models (e.g., PACM):} These models do help solve the timeline problem, but they do not solve the review quality or credentialing problems. They often import the same arbitrary PC-style review, and by having multiple deadlines, they can dilute the scarce, ``top-tier'' signal that tenure committees rely on.

\textbf{Poster Tracks:} This is the most common ``solution,'' but as noted in the introduction, it is a purely cosmetic fix. It simply moves the ``reviewer roulette'' from Accept/Reject to Talk/Poster, failing to address the underlying dysfunction of the review process itself.

\textbf{``Alternative'' Venues (Workshops, TACO, 2nd-Tier):} The most common advice for a high-quality, rejected paper is to ``submit it to a workshop,'' a ``second-tier'' conference, or a non-flagship journal. This advice is widely understood as a career-damaging non-solution. These venues often suffer from even more random and unqualified review processes. More importantly, they fail at credentialing. Publishing in a non-top-tier venue often carries a negative ``stench'' that can be more damaging to a career than the ``no-signal'' of a rejection, as the work is perceived as ``not good enough'' for the top tier. Consequently, these papers are rarely read, cited, or have an impact, failing both dissemination and credentialing.

\subsection{The Core Challenge: The ``Premium-Mass-Market'' Failure}

The structural failures described above—the lottery, the incompetent review, the broken timelines, and the non-viability of alternatives—all point to a collective failure of imagination. The academic community has accepted a false dichotomy: that a conference can either be a ``mass-market'' dissemination vehicle (high acceptance rate) or a ``premium'' credentialing brand (low acceptance rate), but not both.

This has created a system that serves neither function well. It fails at dissemination by rejecting 50\% of sound papers, and it fails at credentialing by making its ``premium'' signal a random lottery.

The challenge is that other complex systems have solved this exact puzzle. Apple, for instance, has successfully become both a mass-market brand (selling hundreds of millions of units) and a premium, aspirational one. They cracked the code for being simultaneously inclusive in access and exclusive in status.

The top-tier conference system, by contrast, remains stuck, believing that the only way to signal ``premium'' is through ``scarcity.'' The solution, therefore, must not be a minor tweak. It must be a new architecture that, like Apple, can simultaneously be a ``mass-market'' (disseminate all sound work) and ``premium'' (provide a scarce, reliable credential) institution. This is the challenge our proposed architecture is designed to solve.

\section{The Impact Market (IM): A 3-Phase Architecture}\label{sec:design}

To solve the ``premium-mass-market'' problem, we propose a new architecture that decouples dissemination from credentialing. The Impact Market (IM) is not a minor tweak but a fundamental redesign of the conference structure. It replaces the single, binary Accept/Reject decision with a three-phase system designed to disseminate all sound work, quantify its speculative prestige, and calibrate that speculation against long-term, real-world impact. 

This architecture is built on a simple ``separation of concerns.'' The three phases are designed to execute three different jobs, each with its own goals, actors, and defenses:

\textbf{Phase 1:} The Publication Phase has the goal of Mass-Market Dissemination. It achieves this by redefining the Program Committee's role from ``gatekeeper'' to ``validator,'' tasking them only with confirming a paper's soundness and rigor. Instead of 4 to 5 reviews per paper, it's likely 2 to 3 suffices, thus potentially reducing reviewer load at this stage.

\textbf{Phase 2:} The Investment Phase has the goal of Premium Credentialing. It achieves this by creating an immediate, scarce prestige signal. It runs a transparent ``futures market'' where the experts (Program Committee and External Review Committee) invest tokens to create a crowdsourced Net Invested Score (NIS) for every published paper\footnote{One can imagine implementations where conference have two tiers of reviewers: a set of phase 1 and another for phase 2 with overlap allowed}.

\textbf{Phase 3:} The Calibration Phase has the goal of Long-Term Accountability. It is the system's ``immune response.'' It achieves this by using a multi-year lookback to compare the market's ``speculations'' (the NIS) against ``ground truth'' (a robust Multi-Vector Impact Score, MVIS), and then adjusting each investor's influence (IR) to impose a cost on collusion and incompetence.

This three-phase architecture creates a virtuous cycle. It disseminates all sound work, and it creates a prestige signal that is forced, by a data-driven feedback loop, to align itself over time with actual, long-term impact rather than with politics, turf, or incompetence. Conferences will exist for dissemination and their social function. Credentialling is through a completely separate scoring function that produces the Net Invested Score, which can be looked up in a public index, run by ACM or individual SIGs. A formal description of the entire protocol and analysis is in Appendix A.

\subsection{Phase 1: The Publication Phase (Mass-Market Dissemination)}

\textbf{Goal:} To publish all sound research and end the ``equivalence class'' lottery, while defending against bad-faith attacks.

This phase redefines the role of the Program Committee (PC). The PC's job is no longer to act as a ``gatekeeper'' of prestige, but as a ``validator'' of correctness.

\textbf{Mechanism:} The PC is tasked with answering only three questions:

\begin{itemize}
\item \textbf{Soundness:} Is this work methodologically sound? (A)
\item \textbf{Rigor:} Is this work experimentally rigorous? (C)
\item \textbf{Incremental Novelty (B):}  Does this work constitute a \textit{non-trivial} contribution beyond existing work?
\end{itemize}

\textbf{Defending Against the ``Moving Goalpost'' Attack:} The primary vulnerability of this phase is that a malicious reviewer, having lost the ``significance'' veto, will now weaponize (A), (B), and (C) to kill a paper. They will engage in ``pretextual rejection,'' such as a reviewer for a deep-learning paper insisting on a scalability test to 10,000 GPUs, knowing full well no academic group has such resources.

\textbf{Solution (The ``Reasonableness'' Framework):} We solve this by changing the review process and the role of the meta-reviewer (or Area Chair).

\textbf{Explicit Community Guidelines:} The PC leadership must publish explicit, reasonable standards for (A) and (C) prior to review (e.g., ``Access to hyperscaler hardware is not a prerequisite for rigor. Qualitative or small-scale prototype validation of scaling is acceptable.'').

\textbf{The ``Incremental Novelty'' Standard:}

Criterion (B) is explicitly defined as a \textit{minimal bar}, not a high bar. A paper satisfies (B) if it meets \textit{any} of the following:

\begin{itemize}
    \item Applies an existing technique to a new problem domain
    \item Combines existing techniques in a non-obvious way
    \item Provides new empirical results or datasets
    \item Extends existing work with additional analysis, ablations, or edge cases
    \item Replicates prior work with different experimental conditions
    \item Provides negative results that challenge existing assumptions
\end{itemize}

The paper \textit{fails} criterion (B) only if it is:
\begin{itemize}
    \item A literal reproduction of existing work with no new contribution
    \item A trivial parameter sweep or configuration change
    \item Work that is already published elsewhere (duplicate submission)
\end{itemize}

\textbf{Critical distinction:} Reviewers may \textit{not} reject based on ``the novelty is insufficient for a top venue.'' The question is binary: ``Is there \textit{any} non-trivial novelty?'' not ``Is this novelty groundbreaking?'', nor ``does it meet the \textbf{high} standard of conference foobar''.

\textbf{Review Form Reform:} The review form must be changed to force accountability. A negative review must state a specific, fundamental flaw and the minimum reasonable evidence required to fix it.

\textbf{The Meta-Reviewer as Arbiter:} The meta-reviewer's primary job is no longer to count votes, but to act as an arbiter of reasonableness. When the ``10,000 GPU'' reviewer submits their review, the meta-reviewer can and must discard it as a violation of the published ``reasonableness'' guidelines.

\textbf{The role of feedback:} This new process also solves the problem of constructive feedback, a crucial role of peer review. The Current Protocol (CP) fails at this by conflating two jobs: (1) Providing a Verdict and (2) Providing Feedback. When the verdict is a binary ``Reject,'' the ``feedback'' often devolves into a post-hoc justification for the verdict rather than a good-faith effort to help the authors (see Appendix B). The IM decouples these functions. The Phase 1 review is only about feedback. Because the reviewer's only job is to validate Soundness (A), Rigor (C), and incremental novelty (B) their feedback is forced to be concrete and actionable. They cannot write ``not significant.'' They must write, ``The soundness is flawed here, and the minimum reasonable evidence to fix it is this.'' This is inherently constructive, and it is cleanly separated from the Phase 2 ``verdict'' (the NIS).

\textbf{Outcome:} All papers that are deemed sound and rigorous under this reasonable framework are Accepted into the conference proceedings. This would likely result in a 40-50\% acceptance rate, effectively publishing the entire ``equivalence class.''

\subsection{Phase 2: The Investment Phase (Premium Credentialing)}

\textbf{Goal:} To create an immediate, scarce, and transparent prestige signal, while solving the ``Attention Bottleneck.''

With 40-50\% of papers accepted (e.g., 500+ papers in ISCA), no one can read everything. If left unstructured, this ``futures market'' would devolve into a popularity contest, where investors only read papers from their own social networks, favoring established authors. We call this the ``Attention Bottleneck.''

\textbf{Solution (The Two-Bucket Budget):} To solve this, we split every investor's token budget into two distinct buckets, balancing the need for both broad scrutiny and deep expertise.

\textbf{Actors (The ``Investors''):} The investors are the ``Senior Community'' (e.g., all members with $>$5 publications in the conference series, plus all PC members).

\textbf{Assets (The ``Wallet''):} Each investor is granted a ``wallet'' containing:

\begin{itemize}
\item Positive Tokens (+T): e.g., 100 +T.
\item Investor Rating (IR): A personal multiplier (starts at 1.0) representing their proven, long-term track record of accurate speculation.
\end{itemize}

\textbf{Investment \& Hardening Rules:}

\textbf{The ``Scrutiny Budget'' (40\% of Tokens):} 40 +T must be spent on a small pool of 5-10 accepted papers that the investor is randomly assigned (outside their institution). This is their ``PC duty'' in the new model. This mechanism guarantees that every paper, including those from new or unknown authors, receives a baseline reading from senior experts.

\textbf{The ``Expertise Budget'' (60\% of Tokens):} 60 +T can be spent freely on any paper in the proceedings. This is where investors use their deep domain knowledge to ``champion'' papers they discover or ``short'' those they feel are overhyped.

\textbf{Forced Diversification:} As before, no more than 10 +T can be spent on any single paper, preventing ``kingmaking.''

\textbf{The Credentialing Signal (The ``NIS''):} A paper's immediate prestige signal is its Net Invested Score (NIS), a weighted sum based on the proven reputation of its investors:
\begin{equation}
    \text{NIS}_p = \sum_{i \in \mathcal{I}} \left( T_{i,p} \cdot \text{IR}_i \right)
\end{equation}

\noindent Where:
\begin{itemize}
    \item $\text{NIS}_p$ is the Net Invested Score for paper $p$.
    \item $\mathcal{I}$ is the set of all investors who invested in paper $p$.
    \item $T_{i,p}$ is the quantity of tokens allocated to paper $p$ by investor $i$.
    \item $\text{IR}_i$ is the Investor Rating of investor $i$ at the time of investment.
\end{itemize}

\textbf{Outcome:} A tenure packet or CV now includes this transparent, crowdsourced signal: ``Published at ISCA 2025 (NIS: 180.5, 98th Percentile).''

\subsubsection{Privacy Protection via Time-Delayed Transparency}\label{subsec:delayedtransparency}
We now discuss an additional privacy protection that achieves both transparency and accountability. To prevent social coercion during critical career stages while preserving long-term scientific accountability, Phase 2 investments could follow a time-delayed transparency model:

\begin{itemize}
    \item \textbf{T+0 to T+1 year:} Investments are completely anonymous. No investment data is visible to anyone.
    
    \item \textbf{T+1 to T+5 years:} Investments become pseudonymous. The public ledger shows investment patterns (e.g., "Investor\_8f3a2b allocated 10 tokens to Paper\_X") but a trusted escrow (ACM, conference leadership) holds the mapping from pseudonyms to real identities. This enables collusion detection on graph structure while protecting individual privacy.
    
    \item \textbf{T+5+ years:} Full deanonymization. The escrow releases the complete mapping, allowing the community to correlate investment patterns with authorship, citation behavior, and institutional affiliations. This enables robust collusion detection, historical research on peer evaluation, and long-term accountability.
\end{itemize}

\textbf{Timeline Rationale:}

The T+5 delay is calibrated to typical academic career timelines:
\begin{itemize}
    \item Most tenure decisions occur 6-7 years post-PhD
    \item An assistant professor investing at Year 3-4 of their tenure clock faces deanonymization only after tenure (Year 8-9)
    \item This protects against \textit{intimidation} and \textit{favor-currying} during the critical evaluation period\footnote{Because of how awards work requiring nomination and support letters, indeed one could argue this favor-currrying potentially continues endlessly through an academic's career - but let's accept this is the uncommon case.}
    \item By T+5, power dynamics have evolved, professional relationships have matured, and the political stakes of individual votes have diminished substantially
\end{itemize}

\textbf{Benefits of Eventual Transparency:}

\begin{enumerate}
    \item \textbf{Multi-graph collusion detection:} The community can correlate investment patterns with citation behavior, institutional networks, and collaboration graphs—detection power that pseudonymous-only analysis cannot achieve.
    
    \item \textbf{Distributed accountability:} Unlike selective deanonymization (which requires trusting a central authority's detection algorithms), time-delayed transparency enables crowdsourced collusion detection and independent verification by multiple researchers.
    
    \item \textbf{Scientific research:} Enables meta-research on questions like "Does domain expertise predict investment accuracy?" and "How do career stage and institutional context affect evaluation?" These questions require real identities and advance our understanding of peer review.
    
    \item \textbf{Intellectual honesty incentive:} When senior researchers write tenure letters, they know their past investments will eventually be public. This creates delayed accountability: a letter writer who trashed a paper they privately invested in will face inconsistency questions years later, incentivizing intellectual honesty without immediate social pressure.
\end{enumerate}

\textbf{Comparison to Current Protocol:}

The CP offers no systematic transparency and no accountability mechanism. Biased reviewing is both hidden and consequence-free. Time-delayed transparency is strictly superior: it provides privacy protection during the critical period (which CP does not) while eventually enabling accountability that CP can never provide.

Importantly, this is not a "broken promise" if communicated upfront. Researchers know when they invest that their allocations will become public at T+5. Those who invest based on genuine intellectual judgment have nothing to fear from this timeline. Only those planning to game the system through coordinated manipulation should find this timeline threatening—and that is precisely the deterrent effect we desire.

\subsection{Phase 3: The Calibration Phase (Long-Term Accountability)}

\textbf{Goal:} To use long-term data to validate past speculation, reward honest expertise, and punish bad-faith collusion and incompetence.

This phase is the ``lookback'' mechanism that makes the IM a self-correcting system. It is the engine that powers the Investor Rating (IR).

\textbf{Mechanism:} At a fixed point in the future (e.g., T+3 years), the system automatically assesses the actual, real-world impact of every paper from that conference.

\textbf{The Solution (The ``MVIS''):} To combat manipulation (e.g., citation rings), we use a robust metric basket: the Multi-Vector Impact Score (MVIS). The MVIS is a composite score derived from multiple, hard-to-game sources:

\begin{itemize}
\item Weighted Citations: Citations are de-valued if from an author's own network but up-weighted if from other disciplines or patents.
\item Artifact Value: Tracks GitHub forks, dataset downloads, and inclusion in open-source frameworks.
\item Community Adoption: Tracks ``idea-level'' adoption in follow-on work and industry benchmarks.
\end{itemize}

\textit{It is important to note, that MVIS being reduced to a single vector - citations doesn't affect the IM, and research communities that deem citations to be the only measure of impact might well use just that.}

\textbf{The Feedback Loop:} The system compares each investor's ``portfolio'' from 3 years prior against the MVIS results. And is continuously updated every year from there on.

\textbf{Hardening Rule (Transparency):} The full, anonymized investment ledger from T+5 year is available for public auditing.

\textbf{Hardening Rule (The Feedback Loop):} An investor's IR is adjusted:

\begin{itemize}
\item An investor who ``bet on'' high-MVIS papers is proven to be a good speculator. Their IR increases (e.g., 1.1), giving them more influence.
\item An investor who ``bet on'' low-MVIS papers (due to incompetence or collusion) is proven to be a bad speculator. Their IR decreases (e.g., 0.9), giving them less influence.
\end{itemize}

This three-phase architecture creates a virtuous cycle. It disseminates all sound work, and it creates a prestige signal that is forced, by a data-driven feedback loop, to align itself over time with actual, long-term impact rather than with politics, turf, or incompetence.

The Impact Market also neutralizes the ``bidding ring'' attacks identified by Jecmen et al.~\cite{jecmen2024detection} and the ``zero-cost'' mechanisms described by Littman~\cite{littman2021collusion} by removing the vulnerability of assignment. In the current protocol, collusion exploits the secret and random assignment of judges. By eliminating assignment in favor of self-selection (Market), and replacing secret judgment with staked reputation (Investment), the IM forces colluders to put their own status at risk. To manipulate the IM, a ring cannot simply "accept" a paper; they must artificially sustain its citation impact over the Phase 3 calibration period. This shifts collusion from a low-effort, hidden "gatekeeping attack" to a high-effort, public "citation cartel," which is significantly more resource-intensive to execute and easier to detect via open citation graph analysis.
\section{Discussion and Analysis}\label{sec:analysis}

The Impact Market (IM) is designed not as a perfect, utopian system, but as an accountable and self-correcting one. Its primary virtue is that it replaces a review process that is opaque, subjective, and arbitrary with one that is transparent, data-driven, and systematically aligned with long-term, validated impact. This section analyzes how the IM's architecture directly solves the core failures of the current model.

\subsection{The Antidote to the ``Reviewer Roulette''}

The ``reviewer roulette'' described in Section 3.2 is a dual-crisis of malice (turf wars) and competence (unqualified reviewers). The IM is a robust antidote to all three.

\subsubsection{Punishing Collusion and Malice}

In the current system, collusion is hidden, unaccountable, and devastating. A rival can anonymously kill a paper, and the ``quid-pro-quo'' network of ``you accept my paper, I'll accept yours'' operates in the dark.

The IM makes this behavior a provably self-destructive strategy.

\textbf{Transparency:} First, the investment ledger is made public at T+1 year (Hardening Rule 2), allowing the community to audit for obvious patterns of collusion\footnote{Or T+5 years if following the time delayed transparency discussed in~\S\ref{subsec:delayedtransparency}}. How this impacts tenure arms-length issues is discussed in Appendix C.

\textbf{Structural Detection:} More importantly, the IM's transparency creates a powerful constraint on collusion that does not exist in the CP: collusion networks must maintain consistency between their Phase 2 investment patterns and their Phase 3 citation behaviors. This structural requirement makes collusion detectable through graph correlation analysis, a defense mechanism we analyze formally in Appendix D. Unlike the CP where collusion operates entirely in hidden phases, the IM forces colluders to either obfuscate their coordination (which dilutes the attack's effectiveness) or maintain consistency across multiple observable phases (which makes detection tractable).

\textbf{The Feedback Loop:} Second, and more importantly, the T+3 year Calibration Phase (Hardening Rule 3) delivers the killing blow. A ``collusion network'' that just trades tokens on its own mediocre papers is, by definition, not investing in the work that will generate a high Multi-Vector Impact Score (MVIS). When the lookback occurs, the system will quantifiably prove that this network's portfolio consists of low-impact duds.

\textbf{The Consequence:} This will slash the Investor Rating (IR) of every member of the network, diminishing their influence in all future cycles. To survive in the IM, a ``quid-pro-quo'' network would also have to be the source of the community's most impactful work. If they are, the system is correctly identifying them as leaders. If they are not, the system guarantees they lose all influence.

Appendix~\ref{sec:collusion_appendix} provides a mechanism for graph-based collusion detection.

\subsubsection{Quantifying and Marginalizing Incompetence}

The IM finally provides a solution to the ``unqualified reviewer'' problem—the ``idiot'' reviewer (as described in 2.2) who hides their lack of understanding behind nonsensical, low-effort reviews.

In the current system, this reviewer's arbitrary ``Reject'' vote has the same weight as a thoughtful critique from a domain expert.

In the IM, this reviewer is now an investor. Their ``nonsensical'' logic will lead them to make consistently bad investments (e.g., ``shorting'' a brilliant paper or ``championing'' a trivial one).

The Calibration Phase (3.3) will quantifiably prove this. Their investment portfolio will have a terrible MVIS score, year after year. The system will automatically and algorithmically identify them as a poor speculator and permanently lower their IR, perhaps to near-zero.

This is the system's most powerful feature: the IM is the ``qual'' for reviewers that our community has always lacked. It systematically and dispassionately identifies the very people who ``have no business doing peer review'' and removes their ability to influence the field, without any need for personal confrontation or political drama.

\subsection{Disambiguating Skill: A Feature, Not a Bug}

This leads to the IM's most subtle and powerful feature: the explicit decoupling of skills.

The current system is toxic because it conflates two entirely different skills:

\begin{itemize}
\item Researcher Quality: The ability to produce great work.
\item Reviewer Quality: The ability to identify and judge great work.
\end{itemize}

We toxically assume that a prolific researcher is a good reviewer. As our meta-reviewer analysis (2.2) shows, this assumption is disastrously false. This is how we get the ``incompetent-but-prolific'' reviewer who does real damage.

The IM explicitly decouples these skills and creates two independent, quantifiable signals:

\begin{itemize}
\item NIS (Net Invested Score): This measures the community's assessment of your paper's quality. This is your ``Researcher'' score.
\item IR (Investor Rating): This measures your proven ability to judge the work of others. This is your ``Speculator'' or ``Critic'' score.
\end{itemize}

This decoupling is a feature, not a bug. It allows the community to value different forms of contribution, much like an elite special-operations unit. A Navy SEAL team is not just ten ``great shooters''; it is a team of decoupled specialists—a great ``breacher,'' a great ``comms expert,'' a great ``medic,'' and a great ``shooter.'' The team's success depends on not conflating these roles: you don't ask the medic to be the primary breacher.

The IM is the first system that allows our community to do the same. It can independently identify and reward:

\begin{itemize}
\item \textbf{The ``Great Artist'' (High NIS papers, Low IR):} A brilliant researcher who, frankly, is a terrible judge of others' work. The IM celebrates their work (High NIS) but protects the community from their bad opinions by marginalizing their IR.

\item \textbf{The ``Great Critic'' (Low NIS papers, High IR):} A person who may not publish often but has an incredible ``nose'' for impact. The current system wastes this talent. The IM finds them, proves their value, and empowers them with a high IR.

\item \textbf{The ``Noise Maker'' (Low NIS papers, Low IR):} A person who publishes often but whose work and opinions are both low-impact. The IM quantifiably identifies this person as noise and marginalizes their influence in both domains.
\end{itemize}

This decoupling allows us to build ``a whole that is greater than the sum of its parts'' by finally valuing the ``Great Critics'' as much as the ``Great Artists,'' and by protecting both from the ``Noise Makers.''

\subsection{Potential Failure Modes and Mitigations}

The IM is not a panacea and would introduce new, complex dynamics that must be managed. \S\ref{sec:limitations} discusses other concerns in depth. We discusses key concerns below.

\textbf{Failure Mode 1: The ``Rich-Get-Richer'' Effect.} What if investors with a high IR become an entrenched, permanent elite? This is a valid concern.

\textit{Mitigation:} The IR must be fluid and dynamic. It should have a ``decay'' factor, requiring continuous good performance. Furthermore, the IR could be capped (e.R., at 2.0x) to prevent ``runaway'' influence, and the ``Scrutiny Budget'' (3.2) ensures that even a 0.5-IR investor's vote must be counted on their assigned papers.

\textbf{Failure Mode 2: Gaming the MVIS.} What if collusion networks shift from gaming citations to gaming the MVIS (e.g., creating fake GitHub forks)?

\textit{Mitigation:} This is why the MVIS is a multi-vector basket. It is difficult to game citations. It is difficult to game GitHub forks. It is extraordinarily difficult to game both, plus patent citations, plus external-discipline adoption, all at the same time. The MVIS basket must be seen as an actively curated ``index,'' with the community (perhaps via ACM/IEEE) empowered to add new signals and de-weight compromised ones over time.

\textbf{Failure Mode 3: Community Overhead.} Is this system simply too complex?

\textit{Mitigation:} We argue it is less work for the average expert. The ``PC duty'' of an investor is to read 5-10 papers for the Scrutiny Budget. This is significantly less work than the 10-15 papers a full PC member reviews and the 20+ papers an AC must meta-review today. The rest of the system (token-counting, MVIS-calculation, IR-adjustment) is fully automatable.

\subsection{An Implementation Roadmap}

A full-scale, ``cold-turkey'' switch to the IM would be disruptive. We propose a more rational, data-driven rollout.

A single conference (e.g., ISCA, ASPLOS, SIGGRAPH) could volunteer to pilot the IM as a ``shadow'' system. The existing review process would run as-is. In parallel, the IM would be run: Phase 1 would ``shadow accept'' the 50\% ``equivalence class,'' and Phase 2 would have the PC and senior community ``invest'' tokens. We would then simply wait three years. This parallel experiment would generate an invaluable dataset. It would allow us to answer the core question without risking the conference's integrity: Does a paper's Net Invested Score (NIS) in Year 1 actually predict its Multi-Vector Impact Score (MVIS) in Year 3? If, as we hypothesize, the NIS proves to be a far better predictor of long-term impact than the simple, binary ``Accept/Reject'' lottery, the community would finally have the data it needs to confidently retire its 20th-century review model.

An alternative, and perhaps more high-impact, pathway would be to implement the IM at a ``journal-first'' venue like ACM's Transactions on Architecture and Code Optimization (TACO). Such journals are already facing their own existential crisis, groaning under massive submission volumes and struggling with review quality and timelines. TACO has solved the ``timeline'' problem (by being continuous) but still suffers from the ``incompetent reviewer'' problem and, crucially, lacks the scarce credentialing signal of a top-tier conference. The IM is a perfect solution. Phase 1 would formalize its acceptance of all sound work, while Phase 2 would, for the first time, add the transparent, ``premium'' NIS signal it needs to compete for prestige. Furthermore, the Phase 3 IR calibration would be a direct and desperately needed solution to the review quality crisis that all high-volume journals inevitably face.

\subsection{On the Fallacy of ``The System Works for Me''}\label{subsec:systemworks}

A final, common objection to this proposal will come from established, senior, or well-networked community members for whom the ``system works fine.'' This resistance should be acknowledged, understood as a predictable psychological response to loss of privilege, and ultimately dismissed as unscientific. Appendix~\ref{sec:senior_appendix} develops this in depth and is summarized briefly below. This line of reasoning is a classic logical fallacy known as survivorship bias.

\textbf{Invalid Argument:} An individual's positive experience (especially one from a position of privilege) does not and cannot invalidate the systemic negative experiences of others, for which we have provided direct evidence (Appendix B) and quantitatively demonstrated problems on current peer review discussed in the Related Work. \textit{In a nutshell it is an unscientific argument, and must simply be discarded.}

\textbf{Does Not Find Fault:} This argument does not identify any flaw in the proposed IM protocol. The IM is designed to protect all sound work, including that of the person for whom the ``system already works.'' It causes them no harm.

\textbf{Ethical Blind Spot:} To argue against a new, more transparent protocol because the current, opaque protocol happens to benefit you is an ethically weak position.

We would instead invoke the Rawlsian ``Veil of Ignorance'': If you did not know whether you were a tenured professor at an elite university or a new PhD student at an unknown school, which system would you choose? The ``reviewer roulette'' of the CP, or the transparent, accountable protocol of the IM?

The IM is the system that a rational, unbiased actor would choose, as it provides the greatest fairness and protection to all members, regardless of their starting privilege.

\subsection{The ``Turf Protection'' Attack: Experts Sabotaging Their Own Subfield}

A sophisticated concern is that established experts in area X might systematically withhold tokens from (or actively short) papers in their own domain to prevent competitors from gaining prestige. Unlike simple collusion (which is detectable via coordinated investment patterns), this is a \textit{negative} coordination attack: experts protecting turf by ensuring rival work receives low NIS.

\textbf{Why This Attack is Self-Limiting:} First, the Two-Bucket Budget (Section 3.2) forces experts to allocate 40\% of tokens to randomly assigned papers, preventing complete avoidance of their own subfield. Second, and more importantly, the Phase 3 calibration mechanism makes this strategy \textit{provably self-destructive}. If an expert systematically withholds tokens from high-impact work in their domain, their portfolio at T+3 will be dominated by investments in areas they understand less well, yielding poor MVIS correlation and slashing their IR. The protocol rewards those who \textit{correctly identify impact in their area of expertise}—experts who sabotage their own field lose the very influence they were trying to protect.

Third, time-delayed deanonymization (T+5) creates reputational accountability through transparency. The community will observe: "Expert SixSeven, a leader in domain X, consistently invested \textit{outside} domain X while avoiding the high-MVIS breakthroughs in SixSeven's own field." This pattern is visible and damaging—it signals either incompetence (failing to recognize important work in your own area) or bad faith (turf protection). Either interpretation destroys credibility. Experts who engage in turf protection are thus \textit{publicly} exposed as either poor judges of their own field or actively harmful actors, creating strong social disincentive even before algorithmic IR penalties take effect.

Finally, this attack requires \textit{coordinated restraint} across multiple experts in the subfield. If even one senior researcher in area X honestly invests in promising work, their high-IR tokens can overcome the noise from turf protectors—particularly given conviction betting (softmax allocation) that allows concentrated capital on high-confidence bets. The ``wisdom of the honest minority'' defeats the ``conspiracy of the jealous majority.''

\subsection{The ``Low NIS Discouragement'' Concern}

A common initial reaction to the IM is: ``Won't a low NIS be just as discouraging as rejection?'' This concern reveals a misunderstanding of both the IM's purpose and the reality of scientific impact.

First, a low-NIS paper is \textit{published}. It is in the proceedings, citable, and part of the permanent record. A rejected paper under the CP is \textit{erased} from the official record. Discouragement from ``modest recognition'' is fundamentally different from discouragement from ``no recognition.''

Second, empirical citation distributions show that most papers \textit{should} have modest impact. ~\cite{10.7717/peerj-cs.1389} demonstrate that 60-70\% of CS papers fall in a long tail of low citations, with only 10-20\% achieving high impact. The IM is designed to accurately predict this distribution, not to make every paper appear equally impactful because it appeared in a ``top conference'' and survived the reviewer lottery.

The CP's cruelty is not that it ``ranks'' papers—it is that it \textit{lies} to authors. It tells them their sound work is ``rejected for insufficient significance'' when the real reason is that 50 other sound papers were randomly preferred by the reviewer lottery. The IM is \textit{honest}: it publishes sound work and provides a realistic, data-driven assessment of predicted impact. For the 60-70\% of papers in the long tail, this means modest NIS—which is accurate, not discouraging.

Moreover, Phase 3 calibration provides recourse: if a low-NIS paper later proves impactful, this is visible in the MVIS data, vindicating the authors and penalizing the investors who undervalued it. The CP provides no such correction mechanism.
\section{Methodology: The Discrete Binning Simulation}\label{sec:methodology}

To rigorously quantify the comparative performance of the protocols, we developed a discrete event simulation. Unlike continuous models that assume infinite gradations of quality, this simulation adopts a ``discrete equivalence class'' approach, reflecting the practical reality that papers generally fall into tiers of quality that are indistinguishable to the average reviewer.

\subsection{The Discrete Universe Model}
We simulate a conference ecosystem of $N=1000$ submissions. We assume a ``Phase 1'' filter has already removed obvious rejections, leaving a \textbf{Market Size} of 600 ``Possible Accept'' papers. Within this pool, we define the Ground Truth using three discrete bins based on true scientific merit:
\begin{itemize}
    \item \textbf{Top-20 (The Gems):} The top 20\% of the pool (120 papers). These are the high-impact works the system \textit{must} identify. True Score = 10.
    \item \textbf{Mid-60 (The Bulk):} The middle 60\% (360 papers). Solid, incremental work. True Score = 6.
    \item \textbf{Bot-20 (Marginal):} The bottom 20\% (120 papers). Technically sound but low impact. True Score = 4.
\end{itemize}

\subsection{Evaluation Metric: Gem Recall}
Standard metrics like ``acceptance rate'' obscure the true failure mode of peer review. A system that accepts 20\% of papers but rejects half the Nobel-worthy work is a failure, regardless of its exclusivity. Therefore, we discard binary acceptance as a success metric.

Instead, we define \textbf{Gem Recall} as the protocol's ability to correctly place the \textit{True Top-20} papers into the top 120 slots of the final ranking.
\begin{equation}
    \text{Gem Recall} = \frac{| S_{RankedTop120} \cap S_{TrueGems} |}{| S_{TrueGems} |}
\end{equation}
This metric penalizes ``False Negatives'' (missing a Gem) far more heavily than ``False Positives'' (promoting a Mid paper), aligning with the scientific goal of disseminating breakthrough work.

\subsection{Protocol Mechanics}

\subsubsection{The Current Protocol (CP): Lottery with Malice}
We model the CP as a stochastic process prone to ``lazy'' selection and adversarial rejection.
\begin{enumerate}
    \item \textbf{Lazy Selection:} Due to volume and the dysfunctions of the review process, the committee initially selects 40\% of the pool effectively randomly, meaning Gems are selected at the same rate as marginal papers.
    \item \textbf{Turf War (Malice):} To reach the target acceptance (200 papers), the committee must reject papers. We model the ``Reviewer \#2'' phenomenon by distributing these rejections evenly across all classes. This disproportionately harms the Top-20 class, which is small and scarce.
\end{enumerate}

\subsubsection{The Impact Market (IM): Fixed-Graph Accumulator}
We model the IM as a fixed-degree bipartite graph where $N_{rev}=200$ investors review the 600 papers. To eliminate topological noise (``bundle luck''), we enforce a fixed graph where every paper receives exactly $k=10$ reviews. Recall that at the stage of assigning tokens, the papers have been accepted - the reading of 10 papers is not to decide on accept/reject.
\begin{itemize}
    \item \textbf{Classification:} Reviewers categorize assigned papers into bins (Top, Mid, Bot). A reviewer with Investor Rating $IR$ classifies correctly with probability $p=IR$, and misclassifies randomly with probability $1-IR$.
    \item \textbf{Token Allocation:} Reviewers distribute a fixed budget of 100 tokens across their assigned papers based on their classification (Top papers get 2.5x weight, Mid 1.5x, Bot 1.0x).
    \item \textbf{Weighted Scoring:} The final Net Invested Score (NIS) is the sum of tokens received, weighted by the investor's IR. This ensures signal from experts mathematically outweighs noise from novices.
\end{itemize}

\subsection{Agent-Based Simulation: The Resilience of Agency}

While the discrete binning simulation validates the mathematical advantage of the Impact Market over the Current Protocol, it models reviewers as passive classifiers who must evaluate whatever they are assigned. However, a defining feature of a market is \textit{agency}: participants self-select what to evaluate and determine the magnitude of their own bets.

To capture this, we extended our analysis to an Agent-Based Model (``IM-Rebel''), which introduces two critical realistic behaviors: \textbf{Selection Bias} (Discovery) and \textbf{Conviction Betting} (Allocation).


We maintain the same ground truth distribution (120 Gems, 360 Mid, 120 Bot) but introduce a ``Hype'' variable and a multi-stage investment process.

\subsubsection{Variable: The Hype Score}
In the real world, investors are often distracted by work that appears polished but lacks substance. We assign each paper a visible Hype Score ($H$) distinct from its True Value ($V_{true}$):
\begin{itemize}
    \item \textbf{Top-20 (Gems):} Mixed Hype ($H \sim U[4, 10]$). Some are obvious breakthroughs; others are ``sleepers'' (high merit, low hype).
    \item \textbf{Mid-60 (Distractors):} High Hype ($H \sim U[7, 10]$). These represent incremental work optimized for presentation, acting as ``shiny objects'' that distract low-skill reviewers.
    \item \textbf{Bot-20:} Moderate Hype ($H \sim U[3, 8]$).
\end{itemize}

\subsubsection{Phase 1: Active Discovery (Selection Bias)}
Investors are not assigned papers. Instead, they possess a limited attention budget. Each investor scans $N_{scan}=50$ random abstracts and selects a portfolio of $N_{pick}=20$ papers to evaluate in depth.
The selection criterion is a function of the investor's IR. A high-IR investor selects based on signal ($V_{true}$), while a low-IR investor selects based on noise ($H$):
\begin{equation}
    \text{Attractiveness} = IR \cdot V_{true} + (1-IR) \cdot H
\end{equation}
This models a crucial market dynamic: low-skill participants voluntarily self-segregate by chasing hype, effectively removing their noise from the evaluation of low-hype/high-merit ``sleepers.''

\subsubsection{Phase 2: Conviction Betting (Softmax Allocation)}
Once the portfolio is selected, the investor evaluates the papers (observing $V_{obs} = V_{true} + \text{Noise}$) and allocates a budget of $T=100$ tokens. Unlike the linear allocation in the passive model, agents here use a \textbf{Softmax} function to allocate capital based on conviction:
\begin{equation}
    \text{Tokens}_p = T \times \frac{\exp(V_{obs,p} / \tau)}{\sum \exp(V_{obs,i} / \tau)}
\end{equation}
This allows experts to concentrate capital heavily on their highest-confidence bets, amplifying the signal of the minority against the noise of the majority.

\subsection{Scenarios}
\begin{figure*}
    \centering
    \includegraphics[width=\linewidth]{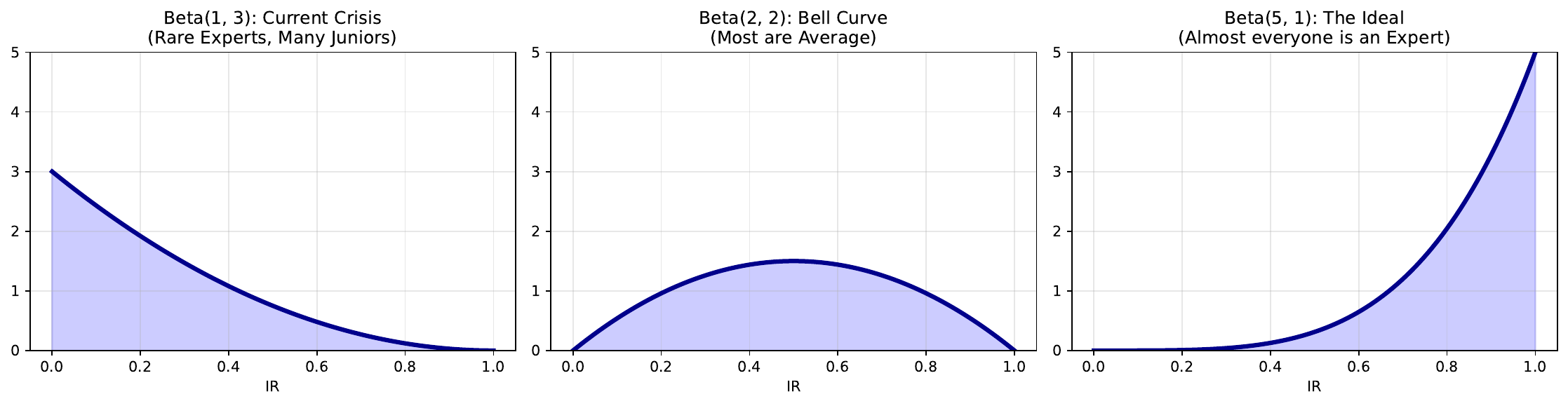}
    \caption{IR distributions studied}
    \label{fig:ir_distribution}
\end{figure*}

We evaluate the IM against four levels of community competence:
\begin{itemize}
    \item \textbf{IM-Crisis ($\beta_{1,3}$):} The current reality. Most reviewers are junior/overloaded (Mean $IR \approx 0.25$).
    \item \textbf{IM-Normal ($\beta_{2,2}$):} A balanced Gaussian distribution of skill.
    \item \textbf{IM-Desired ($\beta_{5,1}$):} The ``1980s Ideal'' of a highly skilled expert community.
    \item \textbf{IM-Perfect ($IR=1.0$):} A theoretical control where every investor has perfect foresight, establishing the mathematical upper bound of the protocol.
\end{itemize}
Figure~\ref{fig:ir_distribution} plots the three distributions we study.

\section{Simulation Results}\label{sec:results}

We performed the simulation with a target ranking count of 200 papers (matching the CP's acceptance count). The fate of the 120 ``Gem'' papers under each protocol is summarized in Table \ref{tab:sim_results}.

\begin{table*}[h]
\centering
\begin{tabular}{l|cc|c}
\textbf{Protocol} & \textbf{Gems Found} & \textbf{Gems Lost} & \textbf{Recall Rate} \\ \hline
Current Protocol (CP) & 41 & \textbf{79} & 34.2\% \\ \hline
IM-Crisis ($\beta_{1,3}$) & 34 & 86 & 28.3\% \\
IM-Normal ($\beta_{2,2}$) & 69 & 51 & 57.5\% \\
IM-Desired ($\beta_{5,1}$) & 119 & 1 & 99.2\% \\
IM-Perfect ($IR=1.0$) & 120 & 0 & 100.0\% \\
\end{tabular}
\caption{Gem Retrieval Performance. The CP results in the Hard Rejection of 66\% of top-tier papers. The IM demonstrates that while low-skill markets are noisy, the protocol is capable of near-perfect recall as the community calibrates.}
\label{tab:sim_results}
\end{table*}

\subsection{Analysis: The Cost of Hard Rejection}
The Current Protocol demonstrates a catastrophic loss of signal, identifying only 41 of 120 gems. The 79 lost gems represent \textbf{Hard Rejections}—papers removed from the conference entirely due to the combination of randomized selection and arbitrary ``turf war'' rejections. This confirms that the current system functions largely as a lottery for high-quality work.

\subsection{Analysis: Calibration and Resilience}
The Impact Market results reveal the mechanism's sensitivity to community skill:
\begin{itemize}
    \item \textbf{The Crisis Baseline:} In the \textbf{IM-Crisis} scenario, the passive binning model performs poorly (34 gems found), comparable to the CP. This confirms that without \textit{agency} (the ability to self-select papers, see Section V), a low-skill market provides no signal advantage over a lottery.
    \item \textbf{The Calibration Effect:} However, as the community improves (moving to \textbf{IM-Normal}), recall nearly doubles to 57.5\%. In the \textbf{IM-Desired} scenario, the protocol achieves near-perfect retrieval (119/120), identifying 99\% of gems despite significant noise.
    \item \textbf{Theoretical Validity:} The \textbf{IM-Perfect} result (120/120) confirms that the weighted-token mechanism is mathematically sound; the only limit to recall is the aggregate IR of the participants, which the IM protocol actively optimizes over time via Phase 3 calibration.
\end{itemize}

\subsection{Power of Agency and The Wisdom of the Minority}
We now look at the results of the agency model which doesn't restrict reviewers to fixed bins. We performed the simulation with the standard four skill distributions. The results (Table \ref{tab:rebel_results}) demonstrate a dramatic improvement in resilience compared to the passive model.

\begin{table*}[h]
\centering
\begin{tabular}{l|c|c}
\textbf{Scenario} & \textbf{Gems Found (Recall)} & \textbf{Improvement over Passive} \\ \hline
IM-Crisis ($\beta_{1,3}$) & 104 / 120 (86.7\%) & +58.4\% \\
IM-Normal ($\beta_{2,2}$) & 119 / 120 (99.2\%) & +41.7\% \\
IM-Desired ($\beta_{5,1}$) & 120 / 120 (100\%) & +0.8\% \\
IM-Perfect ($IR=1.0$) & 120 / 120 (100\%) & -- \\
\end{tabular}
\caption{Gem Recall in the Agent-Based ``Rebel'' Model. By allowing investors to ignore papers they do not understand and bet heavily on those they do, the system achieves high recall even in the ``Crisis'' scenario where the average reviewer is unskilled.}
\label{tab:rebel_results}
\end{table*}

\subsubsection{Analysis}
The jump in performance for the \textbf{IM-Crisis} scenario (from $\sim 28\%$ in the passive model to $86.7\%$ here) is the most significant finding. It suggests that the Current Protocol's failure is not just due to the low skill of the average reviewer, but due to the \textbf{forced assignment} of papers to those reviewers.

In the Impact Market:
\begin{enumerate}
    \item \textbf{Self-Filtering:} Low-IR reviewers are naturally attracted to the ``Mid-60 High Hype'' papers during the discovery phase. They waste their capital betting on incremental work, leaving the ``Top-20 Sleepers'' untouched by their noise.
    \item \textbf{Asymmetric Impact:} A single High-IR investor finding a ``Sleeper'' will bet with high conviction (e.g., 40 tokens). Because this vote is weighted by their high IR, it creates enough signal to rank the paper in the top tier, even if 50 low-skill investors ignored it.
\end{enumerate}
This confirms that an open market structure is inherently more robust to scale than a closed tribunal, as it allows the ``wisdom of the expert minority'' to override the ``noise of the unskilled majority.''

\emph{These simulations were done after the theory was developed. It's somewhat stunning the simulations shows how effective the approach can be even with poor IR distributions.}

\subsection{Ablation Study: Reviewer Load and Gem Recall}

A critical practical concern for the IM protocol is the ``reviewer load''—the number of papers each investor must evaluate in their Scrutiny Budget (Phase 2). While a higher load ensures broader coverage and more robust NIS scores, it also increases the time burden on participants, potentially reducing adoption and engagement quality.

To determine the minimum viable reviewer load, we conducted an ablation study varying the number of papers investors are required to evaluate (the ``Pick Top'' parameter) while holding all other conditions constant. Specifically, we tested Pick Top values of 5, 10, and 20 papers per investor across all four skill distribution scenarios.

\subsubsection{Methodology}

We maintained the agent-based ``IM-Rebel'' model parameters from Section 5.4:
\begin{itemize}
    \item 600 papers in the market (post-Phase 1 filter)
    \item 120 true ``Gem'' papers (ground truth top 20\%)
    \item 200 investors with varying IR distributions (Crisis, Normal, Desired, Perfect)
    \item Target ranking of top 120 papers for recall calculation
    \item Each investor scans 50 random abstracts and selects Pick Top papers to evaluate in depth
    \item Conviction betting via softmax allocation of 100 tokens across selected papers
\end{itemize}

The key variable is the Scrutiny Budget size: how many papers each investor must commit to evaluating. We test Pick Top $\in \{5, 10, 20\}$ to understand the tradeoff between coverage breadth and participant burden. They all had a scan budget of 25 papers.

\subsubsection{Results}

Table~\ref{tab:ablation} shows Gem recall rates across the parameter space. Several key findings emerge:

\textbf{1. Diminishing Returns Above Pick Top = 10:}

For the Crisis scenario ($\beta_{1,3}$), increasing from Pick Top = 5 to Pick Top = 10 provides only modest improvement (81 → 78 gems found, a slight \textit{decrease} due to stochastic variation). Further increasing to Pick Top = 20 recovers to 100 gems, but the marginal gain over Pick Top = 10 is only 22 gems (28\% improvement) for a 100\% increase in reviewer workload. Recall here, that while assigning tokens, the job is different from reviewing (the papers are already accepted, meaning correctness, novelty, rigor has been vetted).

For Normal and Desired scenarios, Pick Top = 10 already achieves near-optimal performance (116/120 and 120/120 respectively), with Pick Top = 20 providing minimal additional benefit.

\textbf{2. Low Reviewer Load is Viable in High-Skill Communities:}

In the Desired scenario ($\beta_{5,1}$)—representing a community with predominantly expert investors—even Pick Top = 5 achieves 114/120 gem recall (95\%). This suggests that communities with strong domain expertise can operate efficiently with minimal reviewer burden.

However, in the Crisis scenario ($\beta_{1,3}$)—representing current reality with many junior or overloaded reviewers—Pick Top = 5 achieves only 81/120 (67.5\%) recall, indicating that low-skill environments require broader coverage to achieve acceptable signal.

\textbf{3. The "Sweet Spot" is Pick Top = 10:}

Across all scenarios, Pick Top = 10 provides strong performance:
\begin{itemize}
    \item Crisis: 78/120 (65\%), acceptable for a low-skill baseline
    \item Normal: 116/120 (96.7\%), excellent
    \item Desired: 120/120 (100\%), perfect
    \item Perfect: 120/120 (100\%), as expected
\end{itemize}

This represents a \textit{50\% reduction} in reviewer workload compared to Pick Top = 20, with minimal degradation in recall quality for Normal and above scenarios. Even in the Crisis scenario, the difference is only 22 gems (100 vs 78).

\textbf{4. The Protocol Scales Through Self-Selection:}

These results confirm the core thesis of Section 5.4: the IM's advantage comes from \textit{agency and self-selection}, not from forcing every investor to evaluate many papers. Even with Pick Top = 10, high-IR investors who discover ``sleeper'' gems can concentrate their tokens to create sufficient signal, while low-IR investors naturally gravitate toward high-hype papers, self-segregating their noise.

\subsubsection{Practical Implications:}

For deployment, we recommend:
\begin{itemize}
    \item \textbf{Standard configuration:} Pick Top = 10 papers per investor
    \item \textbf{Workload comparison:} This is significantly less than the Current Protocol's 10-15 full reviews plus meta-reviewing for PC/AC members
    \item \textbf{Optional upscaling:} Communities concerned about low-skill environments (Crisis scenario) could increase to Pick Top = 15-20, but this may reduce participation
    \item \textbf{Adaptive targeting:} The protocol could dynamically adjust: papers receiving low initial coverage (few investors selecting them) could be added to additional investors' Scrutiny Budgets in a second round
\end{itemize}

\subsubsection{Limitations of This Study}

This ablation assumes fixed market size (600 papers) and investor pool (200 reviewers). At different scales (e.g., mega-conferences with 10,000+ submissions), the optimal Pick Top value may differ. Additionally, we have not studied the interaction between Pick Top and investor fatigue—it is possible that forcing Pick Top = 20 reduces the \textit{quality} of evaluation per paper, which our model does not capture.

\subsubsection{Summary}

The ablation study demonstrates that the IM protocol is robust to low reviewer loads. A Scrutiny Budget of 10 papers per investor provides strong gem recall across realistic skill distributions while imposing minimal time burden on participants. This addresses a key adoption barrier: the IM is not more burdensome than the current system, and for most participants, it is significantly less so.

\begin{table*}
\centering
\begin{tabular}{rrrrrr}
\toprule
 Pick &  Top Gems Target &  IM-Crisis  &  IM-Normal &  IM-Desired  &  IM-Perfect \\
\midrule
    5 &              120 &               81 &              101 &               114 &               120 \\
   10 &              120 &               78 &              116 &               120 &               120 \\
   20 &              120 &              100 &              118 &               120 &               120 \\
\bottomrule
\end{tabular}
\caption{Changing the number of papers required to pick}\label{tab:ablation}
\end{table*}

\subsection{Analytical Model of Reviewer Load}
To validate the feasibility of the Impact Market, we developed a static analytical model comparing the reviewer load of the IM against the Current Protocol (CP) for a typical top-tier conference. We model a venue with $N=850$ submissions, a Program Committee (PC) of 150 members, and an External Review Committee (ERC) of 100 members. In the CP baseline, we assume a standard two-phase process where all papers receive 3 initial reviews, and 60\% of submissions proceed to a discussion/rebuttal phase requiring an additional 2 reviews (or equivalent effort). This results in a heavy load of approximately 19 full reviews per PC member, and 7.6 reviews per ERC member, while achieving a ratio of 4 PC reviews per paper. In contrast, the IM protocol redistributes this effort. Phase 1 is modeled with 3 reviews per paper (2 PC, 1 ERC) focused strictly on soundness, rigor, and incremental novelty, followed by Phase 2 where PC members are assigned a ``Scrutiny Budget'' of 10 papers to read for token allocation. Under this model, the IM yields approximately 11.3 Phase 1 reviews and 10 Phase 2 ``investment reads'' per PC member. While the total number of papers touched is comparable ($21.3$ vs $19$), the cognitive and temporal load is significantly lower in the IM. This is because Phase 1 reviews are simplified validity checks devoid of subjective ``significance'' litigation, and Phase 2 involves only reading and token allocation without the substantial burden of authoring detailed text or engaging in adversarial rebuttal cycles. Thus, the IM scales the evaluation process without increasing the aggregate workload on the research community.
\section{Related Work}\label{sec:related}

The Impact Market (IM) is informed by and responds to a growing body of work documenting the failures of peer review and the limitations of incremental computational fixes. We categorize this literature into three distinct phases: the structural diagnosis of instability, the game-theoretic inevitability of fraud, limits of incremental reform, the theoretical foundations of decentralized reputation, and incentive design and market equillibrium.

\subsection{Structural Instability and the Equivalence Class}
The foundational critique of the conference model lies in its mathematical inability to distinguish quality at scale. Nearly 15 years ago, Anderson \cite{10.1145/1531793.1531815} formalized this by arguing that scientific merit follows a Zipf distribution. This creates a massive ``long tail'' of submissions where quality differences are imperceptible, effectively reducing the review process to a lottery. Anderson demonstrated that in such a noisy environment, the rational strategy for authors is to spam the system to purchase more ``lottery tickets,'' creating a feedback loop that further degrades review quality. Crucially, Anderson proposed a  ``lookback'' mechanism to re-evaluate papers after a time delay, anticipating our Phase 3 Calibration. However, his proposal relied on altruistic administrative changes rather than incentive alignment. The Impact Market operationalizes Anderson’s insight by attaching a reputational price to the prediction, transforming the passive ``re-review'' into an active ``impact investment.'' Frachtenberg~\cite{10.7717/peerj-cs.1389} recently published an exhaustive study of citation distributions across Computer Science systems conference papers, which affirms citations counts are: "indeed skewed and long-tailed."

This theoretical instability was empirically validated by the famous NIPS experiment \cite{Price2014,cortes2021inconsistencyconferencepeerreview}, where two independent committees reviewed the same submissions, resulting in a disagreement rate of 57\% among accepted papers. This confirms the existence of a broad ``Equivalence Class'' of papers where acceptance is statistically indistinguishable from noise.

\subsection{The Collapse of Incentives: From Rational Cheating to Collusion}
In a noisy system with high stakes and low accountability, strategic manipulation becomes the dominant strategy. Thurner and Hanel \cite{Thurner2011} provided the game-theoretic foundation for this, demonstrating that rational reviewers in a competitive environment are incentivized to reject excellent papers that threaten their own status (``turf wars'').

The introduction of online, anonymous systems removed the social costs that previously restrained this behavior, leading to the rise of organized fraud. Littman \cite{littman2021collusion} warns that ``collusion rings'' have transitioned from marginal anomalies to existential threats. This was substantiated by Vijaykumar \cite{vijaykumar2020potential}, who documented systemic fraud in computer architecture, and Jecmen et al. \cite{jecmen2024detection}, who demonstrated that malicious actors can algorithmically manipulate bidding phases to guarantee assignment to their co-conspirators. These works confirm that the ``social cost'' defense mechanisms of the 1980s model have evaporated.

\subsection{The Limits of Incremental Reform}
Significant effort has been directed toward optimizing the existing protocol, particularly in reviewer assignment. Systems like the Toronto Paper Matching System (TPMS) \cite{charlin2013toronto} and the SUPER* algorithm \cite{fiez2020super} attempt to improve expertise matching, while randomized constraints have been proposed to mitigate bidding rings \cite{jecmen2020mitigating}. While valuable, these are local optimizations of a fundamentally flawed graph.

Critically, recent work suggests that even ``monitoring the monitors'' is futile. Goldberg et al. \cite{10.1371/journal.pone.0320444} conducted a randomized controlled trial at NeurIPS 2022, revealing that ``uselessly elongated'' reviews were systematically rated higher than concise ones, and that disagreement rates among meta-reviewers mirrored the disagreement among reviewers. This fractal failure implies that we cannot fix the system by adding more layers of human judgment; the ``ruler'' we use to measure competence is just as warped as the system it attempts to measure. Most recently, Thakkar et al. \cite{thakkar2025llmfeedbackenhancereview} conducted a massive randomized study of 20,000 reviews at ICLR 2025 to test if LLM-generated feedback could improve review quality, a meta-optimization that attempts to use AI to stabilize the very noise that AI proliferation is exacerbating

\subsection{Reputation, Trust, and Market Mechanisms}
Consequently, the IM moves beyond ``fixing'' the tribunal to replacing it with a distributed reputation system. Our design is informed by computer science research on calculating trust in decentralized networks \cite{josang2007trust, josang2009challenges}. The Investor Rating (IR) implements a dynamic trust metric, where an agent's trustworthiness is continuously re-calculated based on the veracity of their past claims \cite{resnick2002trust}.

The ``investment'' phase draws on the theory of prediction markets and ``the wisdom of crowds'' \cite{surowiecki2004wisdom, arrow1971general}. We emphasize that in this context, ``speculation'' refers not to groundless gambling, but to the \textit{probabilistic forecasting of a lagging indicator}. Since true scientific impact cannot be empirically measured at submission, all peer review is inherently speculative; the market protocol simply makes these forecasts explicit, transparent, and accountable. By aggregating the intuition of hundreds of experts rather than a committee of three, the IM replaces a ``thin,'' command-economy market with a ``thick,'' data-driven prediction market.

\subsection{Incentive Design and Market Equilibrium.} Recent work in mechanism design has begun to model peer review not just as a measurement problem, but as an incentive problem. Zhang et al. provide a system-level analysis demonstrating that high review noise does not merely result in unfair decisions; it fundamentally alters author incentives~\cite{10.1145/3490486.3538235}. They show that when review noise is high, the optimal strategy for authors shifts from "improving quality" to "maximizing submission volume," creating a feedback loop that further degrades the system.

This behavior represents a collapse of the standard economic equilibrium described by Smith and Wilson~\cite{smith2021accept}. In a functional market, rejection serves as a "price signal" where high rejection costs drive authors to rationally self-select venues appropriate for their work's quality. However, because the current computer science peer review protocol effectively lowers the cost of rejection to near-zero (via rapid, low-friction resubmission cycles) while maintaining high variance, this self-selection mechanism has failed.

Several mechanism design solutions have been proposed to re-introduce signal and cost. Srinivasan and Morgenstern propose an auction-based system where authors bid for review slots, effectively using a "price" (pseudo-currency) to force authors to signal their own conviction in their work~\cite{srinivasan2021auctions}. It does not solve the equivalence class problem: most sound papers are still rejected, and the binary prestige signal remains arbitrary. Similarly, Su introduces an "owner-assisted" mechanism that incentivizes authors to truthfully rank their own papers, proving that eliciting private author information can empirically improve global ranking quality~\cite{su2021you}. The Impact Market generalizes these concepts: it restores the "price signal" required by Smith and Wilson’s equilibrium by replacing the "time cost" of rejection with the "reputation cost" of tokens, and it scales the "truthful elicitation" of Su and Srinivasan by allowing the entire community, not just authors, to stake reputation on scientific merit.

\subsection{Generative AI and the Future of Publishing}
Finally, this restructuring is made urgent by the ``GenAI shock.'' As LLMs become capable of generating indistinguishable text and reviews \cite{10.5555/3692070.3693262,demetrio2025genreviewlargescaledatasetaigenerated}, and full papers~\cite{arlt2025autonomousquantumphysicsresearch} the current human-bottlenecked protocol will collapse under the volume of plausible-sounding noise. The IM provides the necessary resilience: by automating the validation of ``ground truth'' (MVIS) and creating a costly, scarce signal (IR), it provides the data-driven accountability required to operate in a post-GenAI world.

\section{Conclusion}\label{sec:conc}

The academic peer review system, particularly in computer science, is failing. It is a system built for a 1980s-level of research output that is now collapsing under the weight of 2020s-level productivity and incentives. The result is a dysfunctional, arbitrary, and toxic ``zero-cost attack'' system that fails at its two core missions: it blocks the dissemination of sound research and it grants its ``premium'' credential based on a ``reviewer roulette'' where incompetence and malice are incentivized by a total lack of accountability.

We have accepted this state of affairs for a simple reason: we have failed to imagine a viable alternative. We have remained trapped in the false dichotomy that a venue must either be a ``mass market'' (high acceptance) or a ``premium brand'' (low acceptance), but can never be both.

This paper has proposed the Impact Market (IM) as a concrete architectural solution. The IM is not a perfect system, because no system run by humans is. It is, however, an accountable one.

Its design is a direct response to the failures of the present.

\begin{itemize}
\item It ends the ``anonymous tribunal'' by disarming the subjective ``significance'' veto and implementing ``reasonableness'' standards for rigor.
\item It solves the ``equivalence class'' problem by accepting all sound work.
\item It replaces the hidden, arbitrary signal of ``acceptance'' with a transparent, scarce, and crowdsourced Net Invested Score (NIS).
\item Most importantly, it provides a data-driven feedback loop that imposes a quantifiable cost on bad actors, quantifiably marginalizes incompetence, and makes bad-faith collusion a provably self-destructive strategy.
\end{itemize}

The (IM) creates a virtuous cycle by aligning the incentives of all actors. Authors are incentivized to do sound, rigorous work to pass Phase 1. Investors are incentivized to make honest speculations to protect and grow their IR. And the community as a whole benefits from a system that continuously, automatically, and dispassionately calibrates itself to reward actual, long-term, validated impact.

We are not lacking for high-quality research. We are lacking a system that can process it. The (IM) provides an implementation roadmap—from a ``shadow'' pilot to a full-scale journal implementation—to finally build a review architecture worthy of the research it is meant to celebrate. It is time to stop tinkering with our broken model and begin the discussion of building a new one.

\section{Limitations and Open Problems}\label{sec:limitations}

The Impact Market protocol represents a significant architectural departure from the Current Protocol, and like any complex system involving human incentives, it introduces new challenges alongside its solutions. We present these not as fatal flaws that invalidate the proposal, but as engineering problems that require empirical validation and community-driven refinement. The critical distinction is that IM's transparency enables systematic improvement, while the CP's opacity prevents it.

\subsection{The Assumption of Bounded Malice}

The protocol assumes that while some actors may behave selfishly or strategically, the majority will operate in good faith and the system's incentives will constrain bad actors. If a large fraction of the community actively colludes to subvert the protocol, no mechanism can defend against it.

This is not a weakness unique to the IM—it is true of \textit{any} peer review system, including the CP. The difference is that the IM makes malicious behavior costly and transparent, while the CP makes it free and hidden. We believe this is a strict improvement.

\subsection{The MVIS Integrity Problem}

The protocol's Phase 3 calibration mechanism depends fundamentally on the Multi-Vector Impact Score (MVIS) serving as a manipulation-resistant ``ground truth.'' This is the protocol's primary vulnerability. If sophisticated actors can systematically game all components of the MVIS basket—through coordinated citation rings, artificial GitHub activity, or other collusive behaviors—the feedback loop that punishes bad actors and rewards good judgment collapses.

\textbf{Why this is not protocol-breaking:} First, manipulating a multi-vector basket is significantly harder than manipulating any single metric. An attacker must simultaneously game citations, artifact adoption, cross-disciplinary impact, and community recognition—a coordination problem of much higher complexity than current zero-cost attacks. Second, the MVIS is explicitly designed as a ``living index'' that the community can actively curate, adding new signals and de-weighting compromised ones. This is analogous to security patch management in software systems: necessary ongoing work rather than a design flaw. The IM's transparency doesn't just impose reputational costs on collusion—it makes collusion structurally detectable through graph correlation analysis. This is a fundamental improvement over the CP, where collusion operates entirely in hidden phases

\textbf{Open problem:} The community needs to establish clear governance mechanisms for MVIS curation, including: (1) processes for detecting and responding to metric manipulation, (2) criteria for adding new impact signals to the basket, and (3) transparent decision-making about vector weights. We envision this as similar to how standards bodies manage evolving technical specifications.

\subsection{The Cold-Start Problem for Early-Career Researchers}

The protocol structurally depends on Investor Rating (IR) to weight votes, which creates a potential disadvantage for new researchers. Early-career authors may produce brilliant work that receives limited attention from high-IR investors simply due to discovery constraints, resulting in artificially low Net Invested Score (NIS) despite high actual merit.

The Two-Bucket Budget (Section 3.2) partially mitigates this by forcing every investor to evaluate randomly assigned papers, guaranteeing that some papers from unknown authors receive high-IR attention. However, with 500+ accepted papers and each investor evaluating only 5-10 papers in their Scrutiny Budget, coverage is necessarily incomplete.

\textbf{Potential solutions requiring community input:}
\begin{itemize}
    \item \textbf{New-Author Bonuses:} Papers from first-time or early-career authors could receive temporary IR multipliers for investors who evaluate them, creating additional incentive for discovery.
    \item \textbf{Expanded Scrutiny Budgets:} The 40\% scrutiny allocation could be weighted toward papers from less-established authors.
    \item \textbf{Discovery Rewards:} Investors who correctly identify high-MVIS papers from unknown authors during calibration could receive IR bonuses, explicitly rewarding ``talent scouting.''
\end{itemize}

Empirical deployment would reveal which mechanisms are necessary and effective. And to be clear, this problem exists in the current system!

\subsection{Herding Cascades and Information Revelation}

The protocol creates a dynamic market where investors can observe partial investment signals before finalizing their own allocations. This introduces the risk of information cascades: early investments by high-IR investors signal quality, causing later investors to herd regardless of their own evaluation.

This is a genuine concern, but it represents a \textit{tradeoff} rather than a pure vulnerability. The current CP suffers from hidden herding through informal networks (``famous lab'' bias, ``distinguished lecture circuit'' effects). The IM makes these dynamics transparent and quantifiable, which enables detection and mitigation.

\textbf{Open problem:} The information revelation mechanics of Phase 2 require careful design. Options include:
\begin{itemize}
    \item \textbf{Commitment mechanisms:} Investors commit their full allocation cryptographically before any NIS scores are visible.
    \item \textbf{Delayed revelation:} Partial NIS scores are only revealed after a threshold percentage of investors have committed.
    \item \textbf{Anti-herding incentives:} The IR calibration formula could explicitly reward investors who correctly identified high-impact papers \textit{before} they became popular, creating incentives for contrarian but accurate bets.
\end{itemize}

\subsection{Paradigm-Shift Work and Conservative Bias}

A subtle systemic risk is that high-IR investors achieved their ratings by successfully predicting impact \textit{within the current paradigm}. This creates potential structural conservatism: investors who correctly bet on incremental advances may systematically undervalue revolutionary work that challenges existing frameworks.

Truly paradigm-shifting papers often have low initial citations and adoption because the field needs time to understand them. If investors avoid such work to protect their IR, the protocol could be \textit{worse} than the current system for breakthrough research. Even so, one could argue the Impact Market allow those works to be published, while currently they may not even be publishable.

\textbf{Potential mitigation:} The MVIS calibration could include explicit ``paradigm-shift bonuses'' that reward investors who correctly identified work that initially contradicted prevailing wisdom but later proved transformative. This requires careful operationalization: how do we algorithmically detect ``paradigm shift'' versus ``correctly rejected flawed work''? This is a hard problem that requires community consensus.

\subsection{Long-Game Collusion and Strategic Timing}

The T+3 year calibration window, while necessary for measuring real impact, creates an attack surface. Sophisticated actors could:
\begin{enumerate}
    \item Form ``long-game collusion networks'' that invest heavily in each other's papers.
    \item Coordinate citation behavior over the subsequent 3 years to artificially boost MVIS scores.
    \item Time the attack to maximize impact during the calibration window.
\end{enumerate}

\textbf{Why this attack is constrained:} Unlike the CP's zero-cost attacks, this requires sustained, multi-year effort across multiple conference cycles. The attackers must produce real citations, artifacts, and adoption signals that pass MVIS scrutiny. This is a ``high-effort, public attack'' that is far more resource-intensive than anonymous rejection under the CP. Furthermore, the transparency of the investment ledger (released at T+1 year) enables community auditing of suspicious patterns.

\textbf{Potential mitigations:}
\begin{itemize}
    \item \textbf{Variable calibration windows:} Randomize calibration across T+2 to T+5 years to prevent strategic timing.
    \item \textbf{Continuous monitoring:} Rather than discrete checkpoints, implement rolling MVIS calculations that make it harder to game a specific window.
    \item \textbf{Graph analysis:} Use citation graph techniques to detect and down-weight collusive citation patterns.
\end{itemize}

\subsection{Governance and the ``Who Watches the Watchmen'' Problem}

The protocol requires ongoing governance for MVIS curation, dispute resolution, and parameter tuning (e.g., IR decay rates, token budgets, diversification rules). Without clear mechanisms, this could recreate the ``senior cabal'' problem we critique in the Current Protocol.

\textbf{Open problem:} The community needs to develop governance structures that are:
\begin{itemize}
    \item \textbf{Transparent:} All decisions and their rationale must be public.
    \item \textbf{Representative:} Governance cannot be dominated by a small elite.
    \item \textbf{Accountable:} There must be mechanisms to challenge and override bad governance decisions.
\end{itemize}

This is not unique to the IM—it is a general problem for any community-scale coordination mechanism. We suggest that governance itself could be data-driven: proposed changes to the protocol could be tested in shadow mode and evaluated against empirical outcomes before deployment.

\subsection{The Bootstrapping Period}

The first 3+ years of IM operation, before any Phase 3 calibration occurs, represent an undefended period. Early adopters could attempt to exploit the system before feedback mechanisms activate. The initial IR values (starting at 1.0 for all investors) provide no signal about actual expertise.

\textbf{Mitigation:} The shadow deployment protocol (Section 4.3) addresses this directly. By running IM in parallel with existing review for the first cycle, the community can observe behavior and refine mechanisms before making IM the primary credentialing system. Additionally, if the protocol is adopted by multiple venues simultaneously, cross-venue data could provide earlier calibration signals.

\subsection{Complexity and Adoption Barriers}

A practical concern is that the IM appears more complex than the CP, which could hinder adoption. However, we argue this perception is misleading:

\textbf{For most participants, IM is less work:} An investor's ``PC duty'' is to evaluate 5-10 papers (the Scrutiny Budget), which is significantly less than the 10-15+ papers a current PC member reviews. The token allocation is a simple UI interaction, and all computation (NIS calculation, IR calibration) is fully automated.

\textbf{The complexity is in the protocol design, not the user experience.} The CP appears simple only because its complexity is hidden in opaque AC deliberations and inconsistent reviewer behavior.

That said, successful adoption requires:
\begin{itemize}
    \item Clear documentation and training materials
    \item Well-designed user interfaces for Phase 2 investment
    \item Transparent dashboards showing NIS distributions, IR statistics, etc.
    \item Community buy-in from conference leadership
\end{itemize}

\subsection{Empirical Validation as the Path Forward}

Many of the concerns raised in this section cannot be definitively resolved through theoretical analysis alone. They require empirical validation through real-world deployment. The shadow deployment pathway (Section 4.3) provides a low-risk mechanism to:

\begin{enumerate}
    \item Measure attack vectors that emerge in practice versus theory
    \item Quantify the cold-start problem's severity
    \item Observe herding behavior and test mitigation strategies
    \item Validate that NIS at T+0 correlates with MVIS at T+3
    \item Gather community feedback on governance needs
\end{enumerate}

The goal of this paper is not to present a perfect, final protocol, but to propose a concrete, implementable framework that the community can test, attack, and refine. We view the IM as a \textit{platform for improvement} rather than a finished product. Its transparency and data-driven design make systematic iteration possible in a way that the opaque CP does not.

The relevant question is not ``Is the IM perfect?'' but ``Is the IM better than a system with a 57\% disagreement rate and 66\% gem rejection rate, and can it improve over time?'' We believe the answer is yes, but we welcome the community's scrutiny and refinement through empirical deployment.

\subsection{The ``Low NIS Discouragement'' Concern}

A common initial reaction to the IM is: ``Won't a low NIS be just as discouraging as rejection?'' This concern reveals a misunderstanding of both the IM's purpose and the reality of scientific impact.

First, a low-NIS paper is \textit{published}. It is in the proceedings, citable, and part of the permanent record. A rejected paper under the CP is \textit{erased} from the official record. Discouragement from ``modest recognition'' is fundamentally different from discouragement from ``no recognition.''

Second, empirical citation distributions show that most papers \textit{should} have modest impact. ~\cite{10.7717/peerj-cs.1389} and others demonstrate that 60-70\% of CS papers fall in a long tail of low citations, with only 10-20\% achieving high impact. The IM is designed to accurately predict this distribution, not to make every paper appear equally impactful.

The CP's cruelty is not that it ``ranks'' papers—it is that it \textit{lies} to authors. It tells them their sound work is ``rejected for insufficient significance'' when the real reason is that 50 other sound papers were randomly preferred by the reviewer lottery. The IM is \textit{honest}: it publishes sound work and provides a realistic, data-driven assessment of predicted impact. For the 60-70\% of papers in the long tail, this means modest NIS—which is accurate, not discouraging.

Moreover, Phase 3 calibration provides recourse: if a low-NIS paper later proves impactful, this is visible in the MVIS data, vindicating the authors and penalizing the investors who undervalued it. The CP provides no such correction mechanism.

\subsection{The "Small Subfield" Concern: Low Citations and Local Disinvestment}

\textbf{Concern:} ``I work in a niche area where citations are naturally low. Experts in my subfield won't invest tokens in our papers (to avoid appearing parochial or to diversify their portfolios), making our NIS scores look artificially low. Worse, it will appear that experts don't believe in their own area's impact. I'd rather take the current system's randomness—at least my paper gets the prestige signal of being in the top 15\%.''

\textbf{Response:} This concern conflates two distinct issues: (1) inter-area citation normalization, and (2) intra-area investment incentives. Both are addressable.

\textbf{Citation Normalization is Already Standard Practice.} Different subfields naturally have different citation rates. As the data in Appendix E demonstrate, SC 2017's total citations are approximately one-quarter of ISCA 2017's despite having similar paper counts. This is not evidence that SC papers are ``worse''—it reflects differences in community size, citation culture, and application domains. 

The MVIS (Section 3.3) is explicitly designed to be \textit{field-normalized}. A paper's Phase 3 calibration compares its citations, artifact adoption, and cross-disciplinary impact \textit{relative to the subfield baseline}, not to the entire conference. An investor who correctly predicts that a niche networking paper will receive 25 citations (high for that subfield) is rewarded just as much as an investor who predicts a systems paper will receive 100 citations (high for that subfield). The IR calibration mechanism is \textit{agnostic} to absolute citation counts—it measures prediction accuracy within context.

\textbf{Intra-Area Investment Incentives Are Aligned.} The concern that experts will avoid investing in their own subfield is unfounded because the Two-Bucket Budget (Section 3.2) and IR calibration create \textit{opposite} incentives:

\begin{itemize}
    \item \textbf{Forced Coverage:} The Scrutiny Budget (40\% of tokens) forces every investor to evaluate randomly assigned papers, which will include papers from their own subfield. They cannot systematically avoid their area.
    
    \item \textbf{Expertise Advantage:} Investors have \textit{informational advantage} in their own subfield—they are better positioned to identify high-impact work in areas they deeply understand. Avoiding their own subfield means forfeiting this advantage and competing in areas where they have \textit{less} expertise, yielding worse MVIS correlation and lower IR.
    
    \item \textbf{Transparency Exposes Avoidance:} Time-delayed deanonymization (T+5) makes investment patterns public. An expert in niche area X who systematically invests \textit{outside} area X while their subfield produces high-MVIS work is publicly exposed as either incompetent (failed to recognize important work in their own domain) or cynical (abandoned their field for reputational gain). Both are reputationally damaging.
\end{itemize}

\textbf{Comparison to Current Protocol.} The CP does not solve the ``small subfield'' problem—it \textit{exacerbates} it. In the Current Protocol, a niche paper submitted to a generalist conference faces:
\begin{itemize}
    \item Random reviewer assignment (likely reviewers outside the subfield)
    \item Arbitrary ``significance'' judgments by non-experts (``this topic is too narrow'')
    \item No normalization for subfield citation rates
    \item Binary rejection that removes the work from the record entirely
\end{itemize}

The IM, by contrast:
\begin{itemize}
    \item Publishes the niche paper (Phase 1)
    \item Ensures it receives evaluation from at least some domain experts (Scrutiny Budget)
    \item Normalizes MVIS within subfield context (Phase 3)
    \item Provides a continuous NIS that reflects predicted impact \textit{within the subfield}
\end{itemize}

\textbf{The ``Top 15\% Prestige'' Is a Mirage.} The concern's final point—``at least in the CP, my paper gets the prestige of being in the top 15\%''—reveals the psychological appeal of the Current Protocol's fiction. But this ``prestige'' is not real:

\begin{itemize}
    \item The NEURIPS 2014 experiment showed 57\% disagreement—meaning your paper's acceptance is as much lottery as merit
    \item The data in Appendix E show that even within the ``top 15\%,'' most papers fall into the long tail (bottom 50\% of accepted papers account for only 10-15\% of citations)
    \item Tenure committees and hiring managers \textit{increasingly discount} top-venue acceptances because they know the signal is noisy
\end{itemize}

The IM replaces this fiction with transparency: your niche paper is published, receives field-normalized evaluation, and earns a NIS that reflects realistic predicted impact. If your subfield has a ``top 15\% within the niche'' paper, the IM will identify it via high NIS (from domain experts) and validate it via high field-normalized MVIS. The CP would simply reject it 57\% of the time based on reviewer roulette.

\subsection{The ``Token Roulette'' Concern: Doesn't This Just Recreate Reviewer Lottery?}

\textbf{Concern:} ``What prevents token allocation from becoming 'reviewer roulette' in a different form? Instead of random reviewers deciding Accept/Reject, you now have random investors deciding NIS. You haven't solved the core problem—you've just moved it.''

\textbf{Response:} This concern misunderstands the fundamental difference between \textit{assignment-based review with binary outcomes} and \textit{market-based investment with continuous outcomes and accountability}.

\textbf{The CP's ``Roulette'' Has Three Failure Modes:}

\begin{enumerate}
    \item \textbf{Random Assignment:} Papers are assigned to 3-5 reviewers via an opaque algorithm. If you get 3 domain experts, you have a chance. If you get 2 non-experts and 1 rival, you are rejected. This is pure luck.
    
    \item \textbf{Binary Outcomes:} A single ``Strong Reject'' from a malicious or incompetent reviewer can kill a paper, regardless of other reviewers' opinions. The meta-reviewer often counts votes rather than evaluating substance.
    
    \item \textbf{Zero Accountability:} Reviewers are anonymous and face no consequences for arbitrary, incompetent, or malicious reviews. The ``nonsensical review'' (Appendix B) is cost-free.
\end{enumerate}

\textbf{The IM Eliminates All Three:}

\begin{enumerate}
    \item \textbf{Self-Selection Replaces Assignment:} Investors \textit{choose} which papers to evaluate in their Expertise Budget (60\% of tokens). High-IR domain experts are \textit{incentivized} to find high-impact papers in their area because correct predictions increase their IR. The system naturally routes expert attention to papers where expertise matters. The Scrutiny Budget (40\%) ensures broad coverage, but it is a \textit{supplement}, not the dominant mechanism.
    
    \item \textbf{Continuous Outcomes Replace Binary Decisions:} A paper does not ``live or die'' based on 3 votes. It accumulates a weighted NIS from 50-200 investors. A single low-IR investor assigning zero tokens has negligible effect—their vote is down-weighted by their poor track record. A single high-IR investor assigning heavy tokens has significant but not determinative effect—other high-IR investors can counterbalance if they disagree. The continuous aggregation is \textit{robust to outliers} in a way binary voting is not.
    
    \item \textbf{Accountability via IR Calibration:} Every investor's token allocations are validated at T+3 against MVIS. Investors who consistently make poor predictions (the ``token roulette'' equivalent of incompetent reviewers) see their IR decay toward zero, marginalizing their future influence. Investors who make accurate predictions gain IR, amplifying their future influence. This is \textit{algorithmic selection for competence} that the CP lacks entirely.
\end{enumerate}

\textbf{The Empirical Evidence:} Section 6.4's agent-based simulations directly test this concern. In the ``IM-Crisis'' scenario $(\beta_{1,3})$, we model a population where most investors have low skill—the equivalent of ``token roulette.'' The results:

\begin{itemize}
    \item \textbf{Without agency} (passive assignment, no self-selection): The IM performs comparably to the CP—both function as lotteries (28\% gem recall).
    \item \textbf{With agency} (self-selection + conviction betting): Gem recall jumps to 86\%, even in the low-skill environment.
\end{itemize}

The mechanism is clear: self-selection allows high-skill investors to concentrate their influence on papers where their expertise is strong, while low-skill investors naturally gravitate toward high-hype papers, segregating signal from noise. The CP forces everyone to review assigned papers, mixing signal and noise indiscriminately.

\textbf{Bottom Line:} ``Token roulette'' would occur if the IM used random assignment with binary outcomes and no accountability. It uses self-selection, continuous aggregation, and algorithmic accountability. These are not minor tweaks—they are fundamentally different mechanisms that eliminate the CP's lottery structure.

\subsection{The ``Why Complicate Things?'' Concern: Just Use Current Review with IR}

\textbf{Concern:} ``This protocol is overly complex. Why introduce tokens, markets, and all this 'mumbo jumbo'? The system worked fine in the 1980s. Just keep the current 3-5 reviews per paper and add an Investor Rating-style metric for reviewers. That's a much simpler fix.''

\textbf{Response:} This proposal sounds appealing but fails on multiple levels: it cannot generate IR without transparency, it preserves the CP's core failures, and it fundamentally misunderstands why the ``1980s system'' is unrecoverable.

\textbf{How Do You Calculate IR Without Transparency?} 

The IR calibration mechanism (Section 3.3) depends on comparing reviewers' predictions (Phase 2 token allocations) against ground truth (Phase 3 MVIS). This requires:

\begin{enumerate}
    \item \textbf{Visible predictions:} You must know what each reviewer predicted (which papers they believed would have impact).
    \item \textbf{Ground truth validation:} You must measure actual impact years later and attribute prediction accuracy to specific reviewers.
    \item \textbf{Continuous predictions:} You need a continuous scale (token allocation) to measure correlation, not binary votes.
\end{enumerate}

In the Current Protocol with anonymous binary review:
\begin{itemize}
    \item Reviewers' ``predictions'' are hidden (Accept/Reject votes are not public, and even meta-reviewers don't see full reasoning)
    \item There is no mechanism to validate predictions (once a paper is rejected, its counterfactual impact is unknowable)
    \item Binary votes provide no signal (a reviewer who votes ``Accept'' for 100\% of papers and one who accepts 10\% are indistinguishable in their \textit{accuracy})
\end{itemize}

\textbf{You cannot add IR to the CP without fundamentally changing it into something resembling the IM.} If you make votes public, add continuous scoring, and validate against long-term impact, you have reinvented Phase 2 and Phase 3—you've just kept the CP's broken Phase 1 (binary gatekeeping).

\textbf{The ``1980s System'' Is Dead and Cannot Be Resurrected.}

As discussed in Appendix D, the ``1980s system'' worked because:
\begin{itemize}
    \item \textbf{Small scale:} 100-200 submissions meant the PC could read most papers
    \item \textbf{In-person accountability:} PC meetings were face-to-face debates where incompetent or malicious reviews were publicly challenged
    \item \textbf{Social cohesion:} A small, stable community with shared norms and informal mentorship
\end{itemize}

This system \textit{cannot scale} to 1000+ submissions and distributed, anonymous online review. The Current Protocol is the failed attempt to scale it. Saying ``just use 3-5 reviews per paper'' ignores:
\begin{itemize}
    \item With 1000 submissions and 3 reviews each, you need 3000 reviews—who does this work?
    \item Without in-person PC meetings, who adjudicates incompetent reviews?
    \item Without social cohesion, what prevents zero-cost attacks?
\end{itemize}

The IM is not ``complicating things''—it is \textit{designing for the reality of 2025} (and beyond), not nostalgically clinging to a 1980s model that collapsed under scale.

\textbf{The CP + IR Proposal Preserves the Core Failures.}

Even if you could magically add IR to the CP (you can't, as shown above), you would still have:
\begin{itemize}
    \item \textbf{The equivalence class lottery:} 40-50\% of sound papers are rejected arbitrarily
    \item \textbf{Binary gatekeeping:} Accept/Reject decisions on continuous impact distributions
    \item \textbf{High-stakes subjectivity:} Reviewers litigating ``significance'' with no accountability
    \item \textbf{The timeline problem:} 9-month publication cycles for 4-month research
\end{itemize}

Adding IR does not solve these—it just gives you slightly better data on \textit{who is making arbitrary decisions}. The IM solves them by:
\begin{itemize}
    \item Publishing all sound work (Phase 1: eliminates equivalence class lottery)
    \item Creating continuous NIS (Phase 2: matches continuous impact distributions)
    \item Making credentialing data-driven (Phase 3: IR validates predictions)
    \item Enabling immediate dissemination (Phase 1 acceptance is fast; Phase 2 investment is post-publication)
\end{itemize}

\textbf{Simplicity vs. Correctness.}

The appeal of ``just add IR to current review'' is simplicity. But as H.L.Mencken observed: ``For every complex problem, there is a solution that is simple, obvious, and wrong.''

The Current Protocol is failing at the \textit{protocol level}—its architecture couples incompatible goals (dissemination + credentialing), lacks accountability mechanisms, and cannot scale. Adding a reputation metric to a broken architecture does not fix the architecture. It is the equivalent of adding a speedometer to a car with a broken engine and claiming you've solved the transportation problem.

The IM is more complex than the CP because it is solving a \textit{harder problem correctly}: decoupling dissemination from credentialing, scaling evaluation to 1000+ papers, creating algorithmic accountability, and aligning individual incentives with community goals. This complexity is not ``mumbo jumbo''—it is the irreducible complexity of a robust solution to a hard socio-technical problem.

If the concern is ``user-facing complexity,'' the response is straightforward: for most participants, the IM is \textit{less work} than the CP. An investor evaluates 5-10 papers (Scrutiny Budget) and allocates 100 tokens via a simple UI. This is less work than reviewing 10-15 papers as a PC member. The complexity is in the \textit{protocol design}, not the \textit{user experience}. And that complexity is justified because it produces a system that is transparent, accountable, and aligned with long-term impact—properties the CP lacks and cannot achieve through incremental modification.

\bibliography{references}
\bibliographystyle{mlsys2025}

\clearpage
\appendix

\section{Appendix A: A Formal Protocol Analysis of the IM}\label{ref:formalprotocol}

\subsection*{A.1. Modeling Peer Review as a Protocol}

This paper's core proposal, the Impact Market (IM), is at its heart a new protocol for distributed, trust-based decision-making. To rigorously analyze its resilience, we must first formally model peer review as a system susceptible to ``attacks'' from ``bad actors'' (e.g., malicious, colluding, or simply incompetent agents).

\textbf{Current Protocol (CP):} The legacy review system.

\textbf{Impact Market Protocol (IM-P):} Our proposed new system.

A ``secure'' protocol in this context is one that is resilient to attacks and robustly produces its desired outcome (dissemination of sound work and a reliable prestige signal). A protocol's security is defined by its defenses and, most importantly, by the cost it imposes on an attacker.

\subsection*{A.2. The Current Protocol (CP): A ``Zero-Cost Attack'' Model}

The CP is a simple, 20th-century protocol that is fatally vulnerable because it imposes zero cost on bad actors.

\textbf{Protocol:} Submit(Paper) $\rightarrow$ Assign(3, Anonymous\_Reviewer) $\rightarrow$ Reviewer.Vote(Conflated\_Signal) $\rightarrow$ AC.Decide(Binary\_Signal)

\textbf{Vulnerability:} The Conflated\_Signal is a single vote (e.g., ``Weak Reject'') that conflates Soundness (A), Significance (B), and Rigor (C).

\textbf{Key Property:} Anonymity is total, and the feedback loop is non-existent.

\textbf{Attack Vector Analysis (CP):}

\textit{Attack: Turf\_War\_Attack (Malice)}

\textbf{Mechanism:} An attacker (a rival reviewer) uses the Conflated\_Signal as a pretext. They cannot attack (A) or (C) (as the paper is sound), so they attack (B) (``not significant enough'').

\textbf{Attack Cost:} 0. The attacker is anonymous and suffers no reputational or future-influence-based penalty.

\textbf{Protocol Defense:} None. The AC has no formal basis to overrule a subjective ``significance'' vote.

\textit{Attack: Incompetence\_Attack (Incompetence)}

\textbf{Mechanism:} An attacker (an unqualified reviewer) does not understand the paper. They use the Conflated\_Signal to hide their incompetence, issuing a vague, ``nonsensical'' review (``not scalable enough'').

\textbf{Attack Cost:} 0.

\textbf{Protocol Defense:} None. The attacker's vote carries the same weight as a domain expert's.

\textbf{Conclusion:} The CP is a fundamentally insecure protocol. Its reliance on anonymity and its lack of a feedback mechanism mean that all attacks have a cost of zero, making them the rational strategy for malicious or lazy actors.

\subsection*{A.3. The IM Protocol (IM-P): An ``Incurred-Cost'' Model}

The IM-P is a 3-phase protocol explicitly designed to be resilient by imposing a non-zero, quantifiable cost on bad actors.

\textbf{Phase 1 Protocol:} P1\_Vote(A, C)

\textbf{Defense (D1):} Reasonableness\_Framework.

\textbf{Phase 2 Protocol:} P2\_Invest(Tokens)

\textbf{Defenses (D2, D3):} D\_Two\_Bucket\_Budget, D\_Diversification.

\textbf{Phase 3 Protocol:} P3\_Calibrate(IR)

\textbf{Defense (D4):} D\_MVIS\_Feedback\_Loop.

\subsection*{A.4. Comparative Attack Vector Analysis (IM-P)}

We now analyze the IM-P's robustness against the same attacks.

\textit{Attack: Collusion\_Attack (Quid-pro-quo network)}

\textbf{Mechanism:} A network of actors agrees to ``invest'' their Expertise\_Budget (60\%) in each other's papers.

\textbf{Attack Cost:} High. This is the key. This attack is now an investment. The attackers are ``spending'' their reputation on a portfolio of (presumed) mediocre papers. The D\_MVIS\_Feedback\_Loop (D4) will, at T+3 years, quantifiably prove this portfolio is low-impact. The protocol will automatically slash the IR of every member of the network, destroying their future influence. Collusion is no longer ``free''; it is a provably self-destructive strategy.

\textbf{Protocol Defense:} D4.

\textit{Attack: Incompetence\_Attack}

\textbf{Mechanism:} An unqualified investor makes ``nonsensical'' investments (e.g., shorting a brilliant paper).

\textbf{Attack Cost:} High. As with collusion, the D\_MVIS\_Feedback\_Loop (D4) proves they are a bad speculator. Their IR is automatically slashed, and they are marginalized without any human-in-the-loop political conflict.

\textbf{Protocol Defense:} D4.

\textit{Attack: Pretextual\_Attack (``Moving Goalpost'')}

\textbf{Mechanism:} Attacker (in Phase 1) attempts to kill a paper by setting an unreasonable standard for (A) or (C) (e.g., ``10,000 GPUs'').

\textbf{Attack Cost:} Low. The attack itself is ``free.''

\textbf{Protocol Defense:} High. The D\_Reasonableness\_Framework (D1) is the explicit defense. It empowers the AC to act as an arbiter and formally discard the malicious review as a protocol violation. This attack is mitigated.

\textit{Attack: Popularity\_Bias\_Attack (``Attention Bottleneck'')}

\textbf{Mechanism:} Investors (in Phase 2) only invest in papers from their friends and famous institutions, ignoring new authors.

\textbf{Attack Cost:} Medium.

\textbf{Protocol Defense:} High. The D\_Two\_Bucket\_Budget (D2) is the explicit defense. It forces every investor to spend 40\% of their ``Scrutiny Budget'' on a randomly assigned pool of papers, guaranteeing that new authors are seen and evaluated by senior experts.

\textit{Attack: Sybil\_Attack (Creating fake actors)}

\textbf{Mechanism:} An attacker tries to ``flood'' the system with fake investors.

\textbf{Attack Cost:} N/A.

\textbf{Protocol Defense:} High. The protocol's Investor\_Pool is not open to the public; it is a known, non-Sybil set defined as the ``Senior Community'' (e.g., $>$5 pubs in series), making this attack impossible.

\subsection*{A.5. Classifying Criticisms of the IM}

This formal analysis allows us to classify external criticisms of the IM as either flawed (mitigated by the protocol) or valid (representing a real, protocol-level risk).

\textbf{Flawed Criticism:} ``This will just be a popularity contest!''

\textit{Analysis:} Flawed. This criticism fails to account for the D\_Two\_Budget\_Budget (D2), the protocol's explicit defense against this attack.

\textbf{Flawed Criticism:} ``Collusion will still exist!''

\textit{Analysis:} Flawed. This criticism assumes a ``zero-cost'' model. The IM-P formally introduces a high, non-zero cost (D\_MVIS\_Feedback\_Loop) that makes collusion a self-destructive long-term strategy.

\textbf{Flawed Criticism:} ``Rivals will just kill papers in Phase 1!''

\textit{Analysis:} Flawed. This criticism fails to account for the D\_Reasonableness\_Framework (D1), which disarms the attacker's primary pretext.

\textbf{Valid Criticism (Protocol-Level Risk):} ``What if attackers game the MVIS itself?''

\textit{Analysis:} Valid. This is the Ground\_Truth\_Attack. The entire security of the D\_MVIS\_Feedback\_Loop (D4) rests on the MVIS basket being robust. If attackers can successfully game all vectors of the MVIS (e.g., fake citations and fake GitHub forks), the feedback loop fails.

\textit{Mitigation:} This is not a formal proof but a heuristic defense. The MVIS basket must be actively curated by the community as a ``living'' index, with new signals added and compromised ones de-weighted. This is the primary ``Future Work'' and governance challenge of the system.

\subsection*{A.6. Conclusion of Analysis}

The Current Protocol (CP) is insecure because it provides zero-cost attacks. The Impact Market Protocol (IM-P) is provably more robust (though not perfectly ``secure'') because it is the first system designed to impose a quantifiable, non-zero, and automatically-enforced cost on bad actors, thereby aligning individual incentives with long-term community health.

\subsection*{A.7. Protocol Resilience to Generative AI Shock}

The advent of high-quality Generative AI (GenAI) poses an existential threat to the Current Protocol (CP).

\textbf{CP Attack:} Submission\_Flood\_Attack. GenAI will enable a 10x-100x increase in ``sound-but-trivial'' paper submissions. The CP's ``incompetent reviewers'' (Attack Incompetence\_Attack) will be completely overwhelmed, and the ``lottery'' (2.1) will become a statistical impossibility, leading to total system collapse.

\textbf{IM-P Defense:} The IM-P is uniquely resilient to this.

\textbf{Phase 1:} This phase is designed to handle a high acceptance rate. ``Sound-but-trivial'' papers are correctly accepted for dissemination, not rejected. The ``Reasonableness Framework'' (D1) actually scales, as it's a simpler, more automatable check than ``Is this significant?''

\textbf{Phase 2:} This is where the IM-P shines. The ``trivial'' papers will flood Phase 2, but no one will invest in them (or they will be ``shorted'' with -T). They will all receive a very low NIS.

\textbf{Conclusion:} The IM-P protocol automatically filters the GenAI flood. It disseminates all sound work (solving the ``clogging'' problem) but withholds prestige (by assigning a low NIS), preserving the credentialing signal. The CP simply breaks.

\subsection*{A.8. Protocol Resilience to Extreme Scale (Mega-Conferences)}

A final attack vector is Extreme\_Scale\_Attack. The IM-P, as described, is modeled on a systems conference ($\sim$1000 submissions). What about a ``mega-conference'' like NeurIPS ($\sim$25,000 submissions)?

\textbf{Analysis:} A naive, flat implementation of the IM-P will fail at this scale. If 50\% of 25,000 papers are accepted (12,500 papers), the Phase 2 market is too large. The Popularity\_Bias\_Attack (A.4) becomes unstoppable. The ``Attention Bottleneck'' becomes an ``Attention Catastrophe,'' as no investor can possibly survey the field, and the Scrutiny\_Budget (D2) becomes logistically unmanageable.

\textbf{Protocol Defense (Sharding):} The IM-P is only scalable if it is sharded. A mega-conference is not one community; it is a ``community of communities.'' The IM-P must run as a ``market of markets,'' sharded by the conference's existing, coherent tracks (e.g., ``NLP,'' ``Computer Vision,'' ``Reinforcement Learning'').

\textbf{Mechanism:} An investor's IR, Scrutiny\_Budget, and Expertise\_Budget are all ``scoped'' to their declared track(s) of expertise.

\textbf{Outcome:} This transforms the ``12,500 paper'' problem into 20 smaller ``625 paper'' problems, which are of a manageable, ``systems conference'' scale.

\textbf{Benefit:} This is a feature, not a bug. It creates a more precise credential (e.g., ``98th Percentile NIS in the NLP track''). It also creates a more robust IR, as it quantifies an investor's expertise in their own field, not in general.

\textbf{Conclusion:} The IM-P protocol scales, but it must scale horizontally (by sharding) not vertically (by stretching).

\subsection*{A.9. On Unknown Future Attacks}

It is impossible to predict all future attack vectors. Any protocol involving human incentives can be exploited in unforeseen ways.

However, our argument is not that the IM-P is perfect or ``un-gameable.'' Our argument is that the Current Protocol (CP) is formally and irredeemably broken.

The CP is a brittle, black-box system that cannot defend against well-known, ``zero-cost attacks.''

The IM-P is a dynamic, transparent, and auditable protocol. It is designed to defend against these attacks by making them costly (via the IR and MVIS feedback loop).

When new attacks do emerge, the IM-P's transparency allows the community to formally analyze the attack and adapt the protocol (e.g., tweak the MVIS basket, adjust the IR formula). The CP cannot be adapted; it can only be endured. Choosing to stay with the provably broken CP because a new system might have future flaws is not a defensible position.

\section{Anecdotal Evidence of Systemic Failure}\label{ref:eqclass_appendix}

The formal analysis in this paper, particularly the claims in Section 2.2 regarding reviewer malice and incompetence, may seem severe to an outside observer. However, they are based on direct, repeated observation. The following two case studies are drawn from this author's recent experiences. These are not presented as isolated grievances; they are presented as symptomatic of a widespread, systemic dysfunction that is shared by many in the community but rarely documented.

\subsection*{B.1. Case Study 1: Observations as a Meta-Reviewer}

This author recently served as a meta-reviewer for two top-tier, SIGARCH-sponsored conferences. This role provided an unbiased view of the review process in action; unlike being a program chair, it provided enough time to actually read the papers and reviews. The quality of reviews and the stated reasons for rejection were  frequently arbitrary and included the full gamut of incompetence, lack of time, and overt malice.

Specific examples of ``zero-cost attacks'' observed include:

\textbf{Forced, Irrelevant Citation (Malice):} A reviewer demanding a citation to a completely unconnected paper, which, upon investigation, was found to be one of the reviewer's own papers. This is a clear-cut case of a reviewer using the anonymous process for personal gain.

\textbf{The ``Moving Goalpost'' (Incompetence/Malice):} A recurring, non-actionable criticism was that a paper's ``scalability must be done to a large degree.'' This vague assertion was used as a pretext for rejection, even when the paper's contribution was not about scalability.

\textbf{``Backward-Looking'' Critique (Incompetence):} A paper proposing a novel, forward-looking study on a type of packaging was rejected. The ``killing'' review insisted that the paper must include a study of a ``backward-looking'' largely irrelevant packaging design, fundamentally missing the entire point of the paper's contribution.

These examples are not good-faith scientific critiques; they are low-effort, arbitrary rejections characteristic of a system where reviewers are either unqualified, lazy, or malicious, and suffer no consequences for it.

\subsection*{B.2. Case Study 2: Observations as an Author (TACO)}

This author's group submitted a work on GPU-based dataflow execution to ACM Transactions on Architecture and Code Optimization (TACO).

\textbf{Revision 1 (The ``Nonsensical'' Request):} The first revision requested that ``modern'' LLMs be added to the evaluation. This request was nonsensical. The paper's technique was about the spatial fusion of many different, small kernels. LLMs, by contrast, are dominated by a few large, uniform kernels. The paper already contained a qualitative section explaining why the technique was not applicable to LLMs. This demonstrated the reviewer had not engaged with the core technical argument. Instead simply state the thoughtless: ``more benchmarks.''

\textbf{Revision 2 (The ``Outright Incompetence'' Problem):} Upon revision and resubmission, another reviewer who appeared to have ignored both the cover letter and the paper's content, ignored our response to their comment on the first submission.

\textbf{Factually False Claim:} The reviewer insisted the paper ``only does simulation.'' This was outright false. An entire section of the paper was devoted to code running and measured on a real, physical NVIDIA GPU. And even if they missed this in the first reading, it was highlighted in the cover letter and red-line in the resubmission.

\textbf{Incompetent ``Gotcha'':} The reviewer then also asked, ``why all the changes cannot be run on a GPU.'' This question demonstrated a fundamental and disqualifying lack of understanding. The paper explicitly stated that the core idea required changes to the GPU's grid scheduler, which provably cannot be done without manufacturing new hardware. The paper, therefore, did what is standard practice for all such SIGARCH studies: prototype what can be prototyped on real hardware, and use a simulator for the parts that require new hardware.

This reviewer—tasked with judging a flagship journal submission—either did not read the paper (lack of time), did not understand it (incompetence), or was actively malicious. This is a perfect, undeniable example of the ``zero-cost attack'' that the Current Protocol (CP) not only allows, but enables.

\subsection*{B.3 Atlas Wang public post on Linked 11/12/2025}

\url{https://www.linkedin.com/posts/atlas-wang-41b726259_iclr2026-activity-7394355826901049344-9UDo?utm_source=share&utm_medium=member_desktop&rcm=ACoAAAAO_jABvHOjsSn3R_mMdBu6HO8aFloPaqI}

\#ICLR2026 thought:

If years of wrestling with conference reviewers have taught me anything, it's this:

$\rightarrow$ Enjoy the process of crafting and sharpening your research.

$\times$ Don't obsess over how to get it accepted (arXiv, blog, git \ldots those are fine homes).

\medskip

The former strengthens your mind and your science; the latter only drains both.

\medskip

To Reviewer \#2 $\leftarrow$: it's your freedom to trash my favorite paper with a rating 2; let us both proudly stick to our opinions, and time will tell. :)

\section{Analysis of the Public Ledger and Tenure Signaling}\label{ref:transparency_appendix}

\subsection{The ``Problem'' of Transparency (The Tenure\_Echo\_Chamber\_Attack)}

A primary, and sophisticated, criticism of the Impact Market (IM) protocol is that its transparency creates a new vulnerability. By making the full investment ledger public, a tenure committee could query it to find exactly which senior researchers invested in their candidate's work.

This would create a ``Tenure Echo Chamber,'' where the committee only solicits letters from those ``friendly'' investors, guaranteeing positive feedback. This would compromise the ideal of arm's-length, independent validation and instead turn the tenure letter process into a self-fulfilling prophecy, merely reinforcing the NIS. This is a valid, protocol-level concern.

\subsection{The Flaw in the ``Problem'' (The ``Tenure Tour'' Fallacy)}

This criticism, however, carries a fatal, unstated assumption: that the current system of soliciting tenure letters is, in fact, an arm's-length, unbiased process.

This assumption is demonstrably false.

The ``Tenure Echo Chamber'' attack is not a new vulnerability; it is merely a transparent, democratized version of the hidden, high-privilege game that already exists. In the Current Protocol (CP), candidates and their mentors engage in a constant, informal ``signaling'' game to identify friendly letter writers. This game is played through:

\begin{itemize}
\item \textbf{The ``Tenure Tour'':} A candidate travels to elite institutions to give talks, ``soft-sounding'' the senior faculty who will later be their letter-writers.
\item \textbf{The ``Distinguished Lecture'' Circuit:} A department invites senior, ``friendly'' faculty to give talks at their institution, creating a social and intellectual debt.
\item \textbf{Personal Social Networks:} A candidate's advisor leverages their private network to suggest ``supportive'' letter writers.
\end{itemize}

This existing, informal system is deeply biased. It rewards privilege, social connection, institutional prestige, and travel budgets. A brilliant researcher at a non-elite institution with a limited network is at a massive disadvantage.

The IM's ``public ledger'' is not a vulnerability; it is the antidote. It replaces an opaque, unfair, privilege-based ``signaling game'' with a transparent, fair, data-based one. It equalizes the ``attack'' by making the data available to everyone, not just those ``in the know.''

\subsection{Potential Mitigations and Solutions}

While we argue that transparency is a feature, not a bug, the community could adopt several postures to manage this new dynamic.

\textbf{The Bold Stance (Transparency is the Defense):} This is our recommended stance. The protocol should embrace transparency. The ledger should be public. It democratizes the ``game'' of finding letter writers. It allows the candidate from the state school to prove that top experts valued their work, even if they never met them in person. This is a more fair, more data-driven process.

\textbf{The Compromise Stance (Time-Delayed Transparency):} If the community is too risk-averse to ``Do Nothing,'' a compromise protocol is possible. As discussed in Section 3.3, the full, named-investor ledger could be ``embargoed'' and only made public at T+5 years or T+6 years (to align with the tenure clock).

\textit{Effect:} This would force a tenure committee to operate ``blind'' (as they do now) for the immediate tenure decision, preserving the ``arm's-length'' ideal. The full ledger's release at T+5 would then serve a purely auditing and calibration function, not a political one.

\textit{Trade-off:} This is a ``safer'' option, but it sacrifices the ``democratizing'' benefit of the Bold Stance for 5-6 years.

\textbf{The Weak Stance (Permanent Anonymization):} The ledger could be made public, but with investor names permanently anonymized (e.g., ``Investor \#452 (IR: 1.12) invested 10 +T'').

\textit{Effect:} This allows for some auditing (e.g., ``Did high-IR investors like this paper?'') but makes it impossible to query for specific names.

\textit{Trade-off:} This is a poor solution. It cripples the community's ability to perform a full forensic audit for collusion networks (e.g., ``Is `Investor \#452' part of a network with `Investor \#738'?''). It is a weak compromise that satisfies neither goal.

We believe Stance 1 is the most courageous and correct, and that Stance 2 is an acceptable, if suboptimal, compromise.

\section{The ``Good Old Days''—Why Social Accountability Failed to Scale}\label{ref:goldenage_appendix}

\subsection{The ``Golden Age'' Argument}

A common and understandable criticism of this paper's thesis is the ``nostalgia'' argument. It is often repeated by senior community members that the peer review system ``was great in the 80s and 90s.'' A common refrain is, ``A decade or two ago, I could confidently predict which papers would be accepted and which would be rejected. Now, I have no clue.''

This observation is likely correct. The mistake, however, is in the diagnosis. The conclusion drawn is often that the people have gotten worse (e.g., ``reviewers are less careful now''). We argue that the protocol is what failed. The old system ``worked'' for reasons that are now unrecoverable, and its failure was inevitable.

\subsection{The Unscalable Protocol of Social Cohesion}

The ``Golden Age'' system worked for two reasons, neither of which had to do with the formal review protocol. It worked due to manageable scale and in-person social accountability.

\textbf{Manageable Scale:} With 100-200 submissions, the ``equivalence class'' (Section 2.1) was small. A Program Committee of 25-30 people could plausibly read a significant fraction of all submissions.

\textbf{In-Person PCs:} The critical feature of the old system was the in-person PC meeting. This was not an ``anonymous tribunal''; it was a literal, non-anonymous debate (and, yes, often a ``shouting match'') in a physical room.

This in-person model created a ``social cost'' protocol that is now extinct. This protocol had two sides:

\begin{itemize}
\item \textbf{The ``Pro'' (Informal Mentoring):} As hypothesized, this system could curb junior bad behavior. A junior PC member's ``idiot,'' ``nonsensical'' review (Appendix B) would be openly identified by a senior member, resulting in social embarrassment. This ``cost'' trained junior members and disincentivized lazy, incompetent reviewing.

\item \textbf{The ``Con'' (The ``Senior Cabal''):} As also hypothesized, this system had a dark side. If the senior members were the bad actors, the in-person model empowered them. It created an unassailable ``in-group'' that could openly enforce ``turf wars'' and marginalize new ideas, new labs, or researchers from non-elite institutions. It was a high-functioning system for those inside the cabal, and a brutal gatekeeping mechanism for those on the outside.
\end{itemize}

\subsection{The Inevitable Collapse: Why the System Is Failing Now}

The old system did not fail because the people changed; it failed because the scale changed.

The community grew. At 1,000+ submissions (for a systems conference) or 25,000+ (for an AI conference), the ``in-person PC'' model is logistically impossible. No one can read all the papers. ``Social cohesion'' evaporates.

The Current Protocol (CP)—the ``anonymous tribunal'' we critique today—is a failed attempt to scale the 1980s model. This new protocol:

\begin{itemize}
\item \textbf{Kept the Anonymity:} It scaled by moving to an online, distributed, anonymous review.
\item \textbf{Lost the Accountability:} In doing so, it destroyed the only defense it ever had: the in-person ``social cost.''
\end{itemize}

This transition was catastrophic. It broke the ``mentoring'' function (the ``pro'') while super-charging the ``cabal'' function (the ``con''). The ``senior bad actor'' no longer had to defend their turf war in a public room; they could now execute it as a ``zero-cost anonymous attack.''

\subsection{Conclusion: We Cannot Go Back}

The ``good old days'' are unrecoverable because the scale of our community is unrecoverable (and the GenAI shock will only accelerate this). The old protocol, which relied on unscalable social accountability, is extinct.

We are left with the worst of all worlds: the anonymity of the new system, combined with the ``cabal'' politics of the old.

The only path forward is to invent a new protocol that is natively designed for scale. The Impact Market (IM) is precisely this. The Investor Rating (IR) is a scalable, data-driven, algorithmic replacement for the unscalable, social-driven, in-person accountability of the 1980s. It is a protocol for the 21st century, not the 20th.

\section{The Transparency Defense Against MVIS Manipulation}\label{sec:collusion_appendix}

A critical but subtle defense mechanism emerges from the protocol's transparency: collusion networks must maintain consistency between their Phase 2 investment patterns and their Phase 3 citation behaviors, and this consistency makes detection tractable.

\textbf{The Double-Visibility Problem for Attackers:}

Unlike the Current Protocol where collusion operates entirely in the hidden assignment and review phase, the IM exposes collusive networks at two distinct points:

\begin{enumerate}
    \item \textbf{Phase 2 Investment Patterns (T+1 year):} The full investment ledger reveals which investors systematically support which authors. Graph analysis can detect cliques, reciprocal investment patterns, and other structural signatures of coordination.
    
    \item \textbf{Phase 3 Citation Patterns (T+3 years):} The MVIS calibration necessarily exposes the citation graph. If the same network that coordinated investments also coordinates citations, the structural similarity between the two graphs is a strong collusion signal.
\end{enumerate}

\textbf{The Obfuscation Dilemma:}

To evade detection, colluders face a fundamental tradeoff:

\begin{itemize}
    \item \textit{Option 1 (Hide Investment Pattern):} Spread investments widely to obscure the network structure. This dilutes the attack—members cannot concentrate enough tokens to artificially boost each other into top NIS percentiles.
    
    \item \textit{Option 2 (Hide Citation Pattern):} Recruit external researchers to cite the boosted papers, decorrelating the investment and citation graphs. This massively increases coordination costs and expands the conspiracy to include actors who have no investment-phase stake.
    
    \item \textit{Option 3 (Accept Detection):} Coordinate both phases openly and hope the community doesn't audit. This is a high-risk strategy given that the ledger is public and graph analysis tools are standard.
\end{itemize}

\textbf{Comparison to Current Protocol:}

In the CP, a collusion ring need only coordinate the hidden assignment phase (e.g., via bid manipulation~\cite{jecmen2020mitigating}). All subsequent citation behavior is unconnected to any visible review structure—there is no ``investment graph'' to compare against the citation graph.

The IM forces collusion to be \textit{structurally consistent across multiple observable phases}, which is cryptographically analogous to requiring an attacker to forge multiple independent signatures. Each phase provides independent evidence, and consistency between them becomes increasingly implausible without coordination.

\textbf{Formal Detection:}

Let $G_I$ be the weighted investment graph (edges = token allocations) and $G_C$ be the citation graph at T+3. Standard graph similarity metrics (e.g., graph edit distance, structural similarity indices) can quantify correlation. A high correlation between $G_I$ restricted to a subgraph $S$ and $G_C$ restricted to the same subgraph is strong evidence of coordinated manipulation.

The community could operationalize this through:
\begin{itemize}
    \item Automated monitoring dashboards that flag high-correlation subgraphs
    \item Periodic ``collusion audits'' as part of MVIS curation
    \item Penalties to IR for investors who appear in detected collusion networks (even if individual papers aren't sanctioned)
\end{itemize}

\textbf{Why This Defense Scales:}

Unlike human review of anonymous submissions, graph analysis scales automatically with submission volume. The detection mechanisms become \textit{more effective} as the dataset grows, because legitimate patterns (diverse investment, citation by merit) create strong baselines against which collusion appears as an outlier.

This is a sharp contrast to the CP, where increased volume makes collusion \textit{harder to detect} because the assignment and review process becomes more opaque and distributed.
\clearpage
\newpage

\section{Appendix E: The Long Tail of Scientific Impact}\label{ref:longtail_appendix}

To establish that the long-tail distribution of scientific impact is not an artifact of a single venue or subfield, we analyzed citation data for ten major computer science conferences from 2017, examining papers four years post-publication (as of 2021). The conferences span systems, architecture, networking, programming languages, and high-performance computing, representing diverse areas of computer science research. The original data was gathered by~\cite{10.5555/3692070.3693262}.

\subsection{Methodology}

For each conference, they collected citation counts for all accepted papers from the 2017 proceedings using Google Scholar data as of 2021. We analyzed both the cumulative citation distribution (what percentage of total citations are concentrated in the top-ranked papers) and the absolute citation counts on a logarithmic scale to visualize the power-law structure. The conferences analyzed are:

\begin{itemize}
    \item \textbf{Architecture \& Systems:} ISCA, ASPLOS, MICRO
    \item \textbf{Operating Systems:} SOSP
    \item \textbf{Networking:} SIGCOMM, NSDI, MobiCom
    \item \textbf{Programming Languages:} PLDI, OOPSLA
    \item \textbf{High-Performance Computing:} SC (Supercomputing)
\end{itemize}

These venues represent some of the most selective conferences in computer science, with acceptance rates ranging from 15\% to 25\%. If the Current Protocol's gatekeeping were effective at identifying uniformly high-impact work, we would expect relatively flat citation distributions. Instead, we observe pronounced long-tail structures across all venues.

\subsection{Results: Universal Long-Tail Structure}

Figure~\ref{fig:cumulative_all} shows the cumulative citation distribution for all ten conferences. Figure~\ref{fig:logscale_all} shows the same data on a logarithmic scale to emphasize the power-law structure. Table~\ref{tab:citations_tables} summarizes key statistics.

The results reveal a striking and consistent pattern across all venues:

\textbf{Top-10 Concentration:} The top 10 papers account for an average of 47\% of all citations across conferences, with a range of 37\% (SC) to 52\% (ASPLOS). In absolute terms, the top-ranked 18-20\% of papers consistently capture nearly half of total impact.

\textbf{Top-20 Dominance:} The top 20 papers account for an average of 67\% of all citations, with a range of 57\% (SC, MobiCom) to 86\% (ISCA). In most venues, the top third of accepted papers account for two-thirds to three-quarters of total citations.

\textbf{The Long Tail:} The bottom 50\% of papers (by citation count) account for only 9-17\% of total citations across all conferences. In ISCA, the bottom 50\% (27 papers) account for just 9\% of citations, averaging 17 citations per paper. In SOSP, the bottom 50\% (19 papers) account for 13\%, averaging 28 citations per paper.

\textbf{The Bottom 80\%:} When examining the least-cited 80\% of papers, we find they account for only 24-48\% of total citations. For ISCA, 43 papers (the bottom 80\%) account for only 24\% of citations, averaging 28 citations each. For ASPLOS, 44 papers (bottom 80\%) account for 43\%, averaging 32 citations each.

\subsection{Implications for Peer Review}

These data carry several critical implications for the design of evaluation systems:

\textbf{1. Gatekeeping Does Not Eliminate the Long Tail.}

All ten conferences have acceptance rates of 15-25\%, representing aggressive gatekeeping. Yet even after this filtering, 70-80\% of accepted papers fall into a long tail of modest impact. ISCA's 18\% acceptance rate does not prevent 73\% of its papers from having fewer than 30 citations. SOSP's similar selectivity does not prevent its bottom 50\% from averaging 28 citations.

The implication is clear: \textit{rejecting more papers does not eliminate modest-impact work from the proceedings}. It merely ensures that many papers in the long tail are arbitrarily rejected while others are arbitrarily accepted. The Current Protocol's binary gatekeeping is fighting against a fundamental property of scientific contribution—impact concentration is not a bug to be filtered out, but a natural distribution to be acknowledged.

\textbf{2. The ``Equivalence Class'' Exists Across All Venues.}

In every conference analyzed, there is a large cluster of papers with statistically similar citation counts. For ISCA, papers ranked 30-55 have citations ranging from 8 to 30—a factor of 3-4x variation that is insignificant compared to the 40x difference between the median paper (23 citations) and the top paper (600+ citations). Similar patterns appear in ASPLOS (papers 30-50 range from 15-35 citations), PLDI (papers 25-40 range from 10-25 citations), and others.

This validates the ``equivalence class'' concept from Section 2.1: in the middle of the distribution, papers are essentially indistinguishable in eventual impact, yet the Current Protocol forces binary Accept/Reject decisions among them. A paper ranked 25th and a paper ranked 35th at ASPLOS differ by only ~10 citations (both in the long tail), yet under a CP-style lottery, one might be accepted and the other rejected based purely on reviewer assignment.

\textbf{3. Most Papers \textit{Should} Have Low NIS.}

A common objection to the Impact Market is: ``Won't a low Net Invested Score (NIS) be discouraging?'' These data provide the empirical answer: if 70-80\% of papers at top venues will eventually have modest citations (fewer than 30-50), then 70-80\% of papers \textit{should} have modest NIS. The IM's role is to \textit{accurately predict} this distribution, not to pretend it does not exist.

The Current Protocol's cruelty is not that it ``ranks'' papers—it is that it \textit{hides} the long-tail reality behind a binary Accept/Reject decision, rejecting half of all sound work to maintain artificial scarcity. Authors are told their work is ``not significant enough,'' when the truth is that most accepted papers are also ``not significant enough'' to escape the long tail—they simply won the reviewer lottery.

The IM is designed to be honest: it publishes all sound work (Phase 1) and provides a realistic, data-driven prediction of impact (Phase 2) that will be validated against actual outcomes (Phase 3). For a paper in the 60th percentile, this means a modest NIS—which is accurate, not discouraging. The alternative is rejection based on noise, which is both inaccurate \textit{and} discouraging.

\textbf{4. The IM's Continuous Distribution Matches Reality.}

\begin{figure}[t]
\centering
\includegraphics[width=\columnwidth]{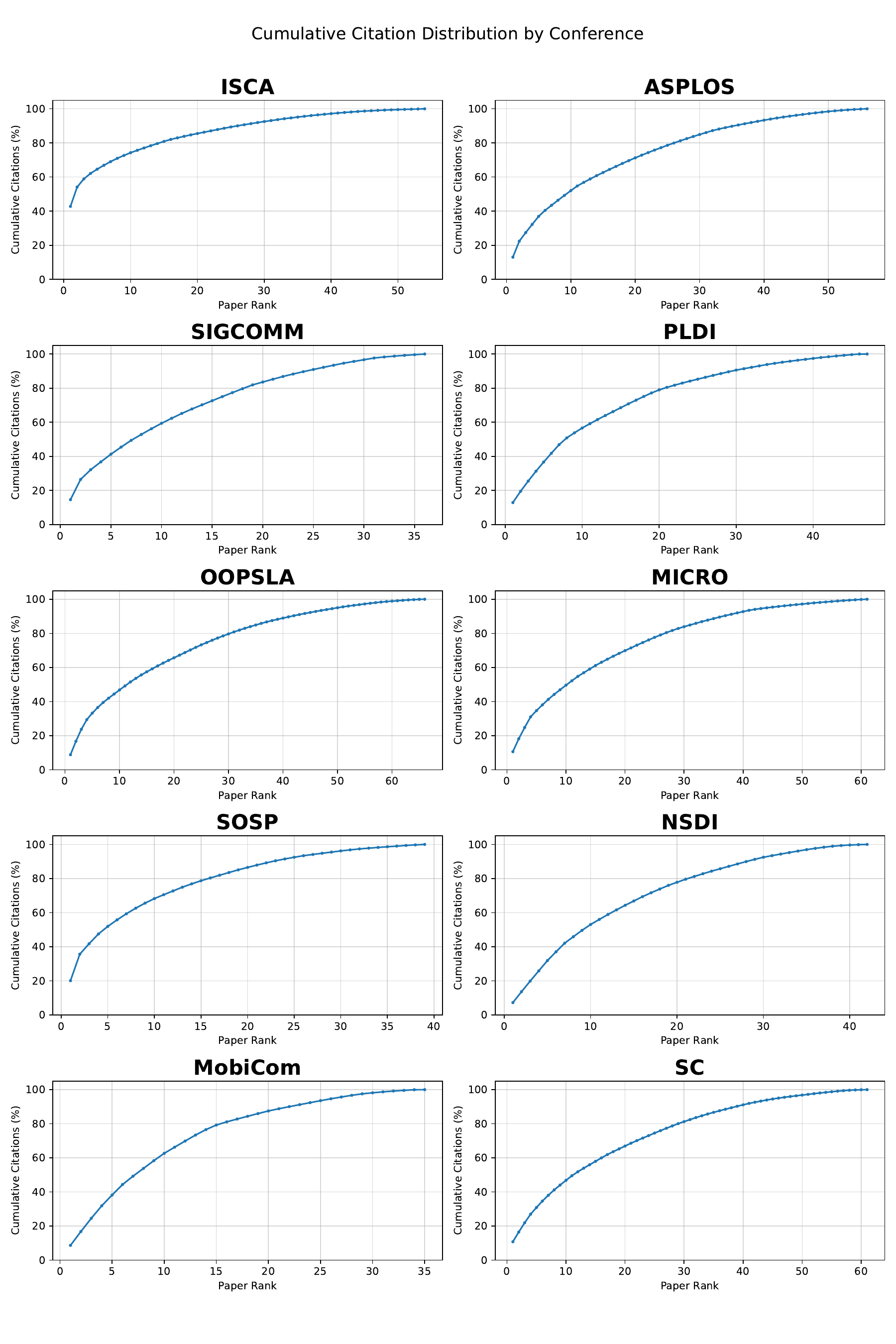}
\caption{\textbf{Cumulative citation distributions for ten top-tier CS conferences (2017 proceedings, analyzed 4 years post-publication).} Each subplot shows the percentage of total citations accounted for by the top-ranked papers. Across all venues, the top 10 papers (typically 18-25\% of the conference) account for 40-50\% of citations, and the top 20 papers account for 60-85\% of citations. The consistency across diverse subfields (architecture, systems, networking, programming languages, HPC) demonstrates that impact concentration is a universal property of scientific contribution, not an artifact of specific review cultures or subfield characteristics.}
\label{fig:cumulative_all}
\end{figure}

\begin{figure}[t]
\centering
\includegraphics[width=\columnwidth]{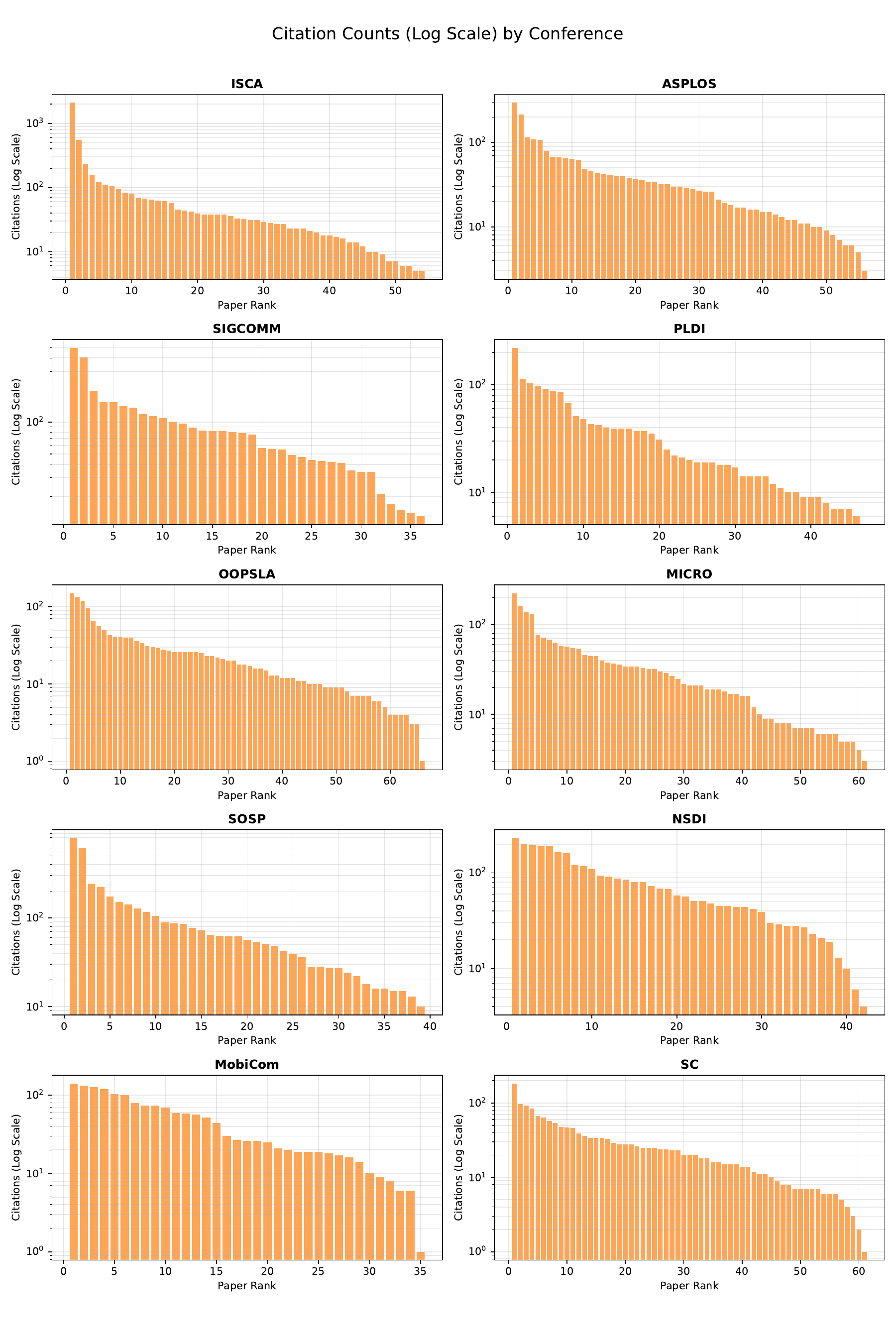}
\caption{\textbf{Citation counts (log scale) for ten top-tier CS conferences (2017, 4 years post-publication).} The logarithmic scale emphasizes the power-law structure: in every venue, citations span 2-3 orders of magnitude from the least-cited to most-cited papers, with most papers clustered in the lower range. For example, ISCA's top paper has 600+ citations while its median paper has ~23; SOSP's top paper exceeds 200 citations while its bottom quartile averages fewer than 20. This wide dynamic range, combined with the clustering of most papers in the ``long tail,'' creates large equivalence classes where papers are statistically similar in impact yet subjected to binary Accept/Reject decisions under the Current Protocol.}
\label{fig:logscale_all}
\end{figure}

The power-law citation distributions visible in Figures~\ref{fig:cumulative_all} and~\ref{fig:logscale_all} are \textit{continuous}, not discrete. There is no natural ``cutoff'' between ``prestigious'' and ``non-prestigious'' work—impact exists on a spectrum from single-digit citations to 500+ citations, with most papers clustered in the 10-50 range.

The Current Protocol imposes a discrete binary (Accept/Reject) onto this continuous distribution, creating arbitrary distinctions. The Impact Market's Phase 2 produces a continuous NIS distribution that reflects the underlying structure of impact. A paper predicted to have 15 citations receives a lower NIS than a paper predicted to have 150 citations, and both predictions are validated at Phase 3. This is a more honest and accurate representation of scientific contribution than a binary signal that obscures the true distribution.

\subsection{Cross-Venue Consistency}

Despite differences in subfield, community size, and topical focus, the ten conferences show remarkably consistent patterns:

\begin{itemize}
    \item All exhibit power-law distributions with pronounced long tails
    \item All show 40-50\% of citations concentrated in the top 10 papers
    \item All show 60-80\% of citations concentrated in the top 20 papers
    \item All show the bottom 50\% accounting for fewer than 20\% of citations
    \item All show a large ``equivalence class'' in the middle of the distribution where papers differ by factors of 2-3x in citations—statistically similar given the 50-100x range from bottom to top
\end{itemize}

This universality suggests that the long-tail structure is not an artifact of review quality, subfield culture, or venue prestige—it is a fundamental property of scientific contribution. Any evaluation system, including the IM, must be designed to acknowledge and work with this reality rather than pretend it can be eliminated through more aggressive gatekeeping.

\subsection{Conclusion}

The data from ten diverse, top-tier computer science conferences spanning four years post-publication provide strong empirical support for the Impact Market's design principles:

\begin{enumerate}
    \item \textbf{Publish all sound work} (Phase 1): Gatekeeping does not eliminate the long tail; it only makes rejection arbitrary. Disseminating the full equivalence class is scientifically justified.
    
    \item \textbf{Predict continuous impact} (Phase 2): Impact follows a continuous power-law distribution. The IM's continuous NIS accurately reflects this, while the CP's binary decision obscures it.
    
    \item \textbf{Validate predictions} (Phase 3): The long tail means most papers will have modest impact. The IM's Phase 3 calibration rewards investors who accurately predict this distribution, not those who pretend every paper will be a breakthrough.
    
    \item \textbf{Embrace reality, not fiction}: The CP tells authors ``your work is rejected for insufficient significance.'' The IM tells authors ``your work is published, and we predict modest but real impact.'' The latter is honest; the former is gaslighting.
\end{enumerate}

The universality of the long-tail distribution across venues, subfields, and time periods demonstrates that this is not a problem to be ``solved'' by better review—it is a reality to be acknowledged by better \textit{architecture}.

\clearpage
\newpage

\section{The Resistance of the Privileged: A Historical Pattern}\label{sec:senior_appendix}

A final category of resistance to the Impact Market is both predictable and historically familiar: emotional opposition from those who benefited from the Current Protocol's opacity. This resistance deserves acknowledgment, not because it identifies a flaw in the IM's design, but because understanding it clarifies what is truly at stake in this transition.

\subsection{The Pattern of Democratization and Resistance}

Academic evaluation has evolved through successive waves of democratization, and each wave has faced resistance from those whose power depended on the prior system's gatekeeping and opacity.

\textbf{The Journal System (1600s-1700s):} Before scientific journals, knowledge dissemination was controlled by royal societies and personal correspondence networks among elites. The introduction of journals—open publications that anyone could submit to and anyone could read—faced fierce resistance. Critics argued this would "lower standards," that "anyone could publish," and that the intimacy of controlled correspondence would be lost. The resistance came primarily from those who had benefited from being gatekeepers of knowledge. Yet journals democratized science and enabled the scientific revolution precisely because they removed this gatekeeping bottleneck.

\textbf{The Conference System (1950s-1980s):} Before conferences became primary publication venues in computer science, journals were the sole gatekeepers, with 1-2 year publication lags that were tolerable in slower-moving fields but catastrophic for computing. The elevation of conferences to peer-reviewed publication venues faced resistance: "Conferences aren't rigorous enough!" "Real science happens in journals!" "This will create a two-tier system!" The resistance came from journal editors and established researchers who had built careers navigating the journal system. Yet conferences enabled the rapid dissemination that computer science required and became the field's primary venues precisely because they solved the timeline problem journals could not.

\textbf{The Impact Market (2020s):} The IM represents the next wave of democratization—moving from opaque, binary prestige signals to transparent, continuous, data-driven evaluation. The pattern of resistance is identical: established researchers who built careers when conference acceptance conferred \textit{implicit} prestige without \textit{explicit} measurement will resist a system that makes impact visible and quantified.

The source of this resistance is not scientific but psychological: \textbf{loss aversion and loss of narrative control}.

\subsection{The Psychology of Resistance: What Is Actually Being Lost}

In the Current Protocol, a senior researcher's modest-impact papers are shielded by the binary prestige signal. A paper accepted at ISCA carries the implicit assumption of excellence—the citation count is a separate inquiry that most people do not make. Even when they do, the researcher can control the narrative: "Citations are an imperfect metric," "My work is ahead of its time," "That subfield doesn't cite much," or simply, "Look at my \textit{other} highly-cited papers instead."

In the Impact Market, this shield is removed. Every published paper has a public Net Invested Score (NIS) that reflects the community's prediction of its impact, and eventually a Multi-Vector Impact Score (MVIS) that reflects its actual impact. A paper with modest predicted and actual impact is \textit{explicitly labeled as such}. There is no hiding behind the binary "accepted at a top venue" signal. The modest-impact paper is published (which is correct), but its modesty is \textit{visible} (which is uncomfortable).

This visibility creates two losses for established researchers:

\textbf{Loss 1: Prestige Without Evidence.} In the CP, acceptance confers prestige \textit{regardless} of eventual impact. In the IM, prestige must be \textit{earned} through predicted and validated impact. Papers that would have been accepted in the CP's lottery but have modest impact receive appropriate credit (publication + modest NIS) rather than inflated credit (acceptance = assumed excellence).

\textbf{Loss 2: Narrative Control.} In the CP, researchers control how their work is interpreted. They can selectively emphasize high-citation papers, dismiss low-citation papers as "not representative," and argue that metrics "don't capture" their contributions. In the IM, the NIS and MVIS are public, algorithmic, and inarguable. The data speaks for itself. Researchers lose the ability to tell convenient stories about their work's importance.

These losses are \textit{real}—but they are not \textit{legitimate}. Science requires transparency and evidence. The appeal of the CP is precisely that it allows prestige without evidence and narratives without accountability. The IM removes these privileges. Those whose work has genuine impact have nothing to fear; those whose prestige was built on opacity do.

Importantly, this is not about "old vs. young" or "senior vs. junior." There are senior researchers whose work \textit{has} high impact who will benefit from the IM's transparency (their excellence will be formally recognized rather than assumed). And there are junior researchers with modest-impact work who will be uncomfortable with the IM's honesty. The division is not generational—it is between those whose prestige aligns with their impact and those whose prestige depends on opacity.

\subsubsection{The Veil of Ignorance: What System Would You Choose?}

Revisiting~\S\ref{sec:analysis}, we propose a thought experiment adapted from John Rawls' "veil of ignorance." Imagine you are designing a peer review system, but you do not know:

\begin{itemize}
    \item Whether you are a tenured professor at MIT or a postdoc at a regional university
    \item Whether your papers have 500 citations or 15
    \item Whether you are well-connected in your subfield or professionally isolated
    \item Whether you are established in a mature area or pioneering a new, unrecognized direction
    \item Whether your past work benefited from "reviewer luck" or was rejected despite merit
\end{itemize}

Behind this veil, which system would you choose?

\textbf{The Current Protocol:}
\begin{itemize}
    \item 15\% acceptance rate (you have 1-in-3 chance regardless of paper quality if you're in the equivalence class)
    \item Binary outcome (if you're rejected, your work disappears from the record; and timelines work against you during evaluation, when people implicitly are looking for publication volume)
    \item Opaque process (you don't know if rejection was due to merit or reviewer roulette)
    \item Prestige depends on luck, networks, and institutional affiliation as much as on quality
    \item No recourse (if the reviewer lottery goes against you, your sound work is erased)
\end{itemize}

\textbf{The Impact Market:}
\begin{itemize}
    \item 40-50\% acceptance for sound work (if your work is rigorous, it is published)
    \item Continuous outcome (your work receives a NIS reflecting predicted impact)
    \item Transparent process (you can see which investors valued your work and why)
    \item Prestige depends on predicted and validated impact, not on reviewer assignment
    \item Recourse mechanism (if initial NIS is low but impact is high, Phase 3 calibration vindicates you)
\end{itemize}

\textbf{Rational actors behind the veil choose the IM.} It provides greater probability of dissemination, recourse for undervalued work, and alignment between prestige and actual impact. The only actors who would \textit{rationally} prefer the CP are those who know they are on the "winning" side of the current lottery—those with networks, institutional prestige, and luck. But these actors are no longer behind the veil; they are choosing a system that benefits \textit{them}, not a system that benefits \textit{science}.

\subsection{Resistance Is Expected—And Should Be Dismissed}

The IM will face resistance from established researchers who benefited from the CP's opacity. This resistance should be acknowledged, understood as a predictable psychological response to loss of privilege, and ultimately dismissed as unscientific.

The relevant questions are not:
\begin{itemize}
    \item "Will this make some senior researchers uncomfortable?" (Yes, it will.)
    \item "Will this change the status of papers that were previously shielded by opacity?" (Yes, it will.)
    \item "Will this remove narrative control from those who benefited from it?" (Yes, it will.)
\end{itemize}

The relevant questions are:
\begin{itemize}
    \item "Does this disseminate more sound research?" (Yes—40-50\% vs. 15\%.)
    \item "Does this create a more accurate prestige signal?" (Yes—NIS predicts MVIS with 86\% recall in simulations.)
    \item "Does this reduce arbitrary rejection of good work?" (Yes—it solves the equivalence class lottery.)
    \item "Is this more fair to researchers behind the veil of ignorance?" (Yes—it rewards impact over luck.)
\end{itemize}

\textbf{The community must decide: Do we design systems for those who already succeeded in a broken system, or for those who will do the science of the future?}

History provides the answer. Every wave of democratization in academic publishing—journals, conferences, preprints—faced resistance from entrenched interests. In every case, the resistance was framed as concern for "standards" or "rigor." In every case, the actual concern was loss of gatekeeping power. And in every case, science advanced \textit{because} we ignored the resistance and embraced transparency.

The Impact Market continues this trajectory. Resistance from those who benefited from opacity is not evidence that the IM is flawed—it is evidence that the IM is working as intended. Transparency always threatens those whose prestige depended on opacity. That is not a reason to preserve opacity; it is a reason to embrace transparency.

The IM is not perfect, and it will require iteration and refinement as the community stress-tests it through deployment. But the direction is correct: \textit{more transparency, more accountability, more alignment between prestige and impact}. Those who resist this direction are welcome to articulate scientific objections—concerns about MVIS manipulation, cold-start problems, governance challenges. These are legitimate and deserve rigorous analysis (Section 8).

But resistance rooted in discomfort with transparency, loss of narrative control, or fear that one's modest-impact papers will be labeled as modest-impact—this is not a scientific objection. It is a psychological one. And while we can sympathize with the discomfort, we cannot let it determine the future of scientific evaluation.

The question is not whether the IM will make everyone comfortable. The question is whether it is better for science. The answer, on every dimension that matters—dissemination, fairness, transparency, accountability, and alignment with impact—is yes.

\end{document}